# Estimation of daily streamflow from multiple donor catchments with Graphical Lasso

**German A. Villalba[1], Xu Liang[1], and Yao Liang[2]**

[1] Department of Civil and Environmental Eng., University of Pittsburgh, Pittsburgh, PA, USA.

[2] Department of Computer and Information Science, Indiana Univ.-Purdue Univ. Indianapolis, IN, USA.

Corresponding author: Xu Liang (xuliang@pitt.edu)

**Key Points:**

- A novel algorithm is presented to select multiple donor gauges for inferring daily streamflow at other locations

- Graphical Lasso method is effective in identifying essential connections among streamflow stations

- The new method shows superior results comparing to existing available methods

**Abstract**

A novel algorithm is introduced to improve estimations of daily streamflow time series at sites with incomplete records based on the concept of conditional independence in graphical models. The goal is to fill in gaps of historical data or extend records at streamflow stations no longer in operation or even estimate streamflow at ungauged locations. This is achieved by first selecting relevant stations in the hydrometric network as reference (donor) stations and then using them to infer the missing data. The selection process transforms fully connected streamflow stations in the hydrometric network into a sparsely connected network represented by a precision matrix using a Gaussian graphical model. The underlying graph encodes conditional independence conditions which allow determination of an optimum set of reference stations from the fully connected hydrometric network for a study area. The sparsity of the precision matrix is imposed by using the Graphical Lasso algorithm with an L1-norm regularization parameter and a thresholding parameter. The two parameters are determined by a multi-objective optimization process. In addition, an algorithm based on the conditional independence concept is presented to allow a removal of gauges with the least loss of information. Our approaches are illustrated with daily streamflow data from a hydrometric network of 34 gauges between 1 January 1950 and 31 December 1980 over the Ohio River basin. Our results show that the use of conditional independence conditions can lead to more accurate streamflow estimates than the widely used approaches which are based on either distance or pair-wise correlation.

**1. Introduction**

Continuous daily streamflow time series are important for a wide variety of applications in hydrology and water resources. Such applications include water supply management, hydropower development, flood and drought control, forecasting of agricultural yield, ecological flow assessment, navigation, rainfall runoff model calibration, design of engineering structures such as highways and reservoirs, and many others (Archfield & Vogel, 2010; Farmer & Vogel, 2013; Parada & Liang, 2010; Razavi et al., 2013; Shu & Ouarda, 2012). However, continuous streamflow data are not available oftentimes due to either no existing streamflow gauges or data gaps in the recorded time series at gauged stations (Huang & Yang, 1998)). Also, data gaps of different time periods exist at different gauged locations within a large river basin (Hughes & Smakhtin, 1996). Furthermore, there is an increasing decline in the hydrometric network density worldwide (Mishra & Coulibaly, 2009; Samuel et al., 2011). For example, the U.S. Geological Survey (USGS) is in the process of discontinuing operations of some streamflow stations nationwide due to budget cuts (USGS, 2019) which has been a serious concern (Lanfear & Hirsch, 1999; Stokstad, 2001; Vorosmarty et al., 2001; Witze, 2013). Therefore, it is critical to develop an effective and general method to fill in data gaps, extend data records of those that have been or will be shut down, and even estimate data for ungauged locations.

The estimation of continuous daily streamflow time series techniques at ungauged or poorly gauged locations can be classified into two broad categories: (1) hydrologic model–dependent methods and (2) hydrologic model–independent methods (Razavi et al., 2013). The latter methods are also called statistical methods (Loukas & Vasiliades, 2014) or hydrostatistical methods (Farmer & Vogel, 2013). Work related to the first category is abundant but has its limitations; and the work related to the second category is relatively limited (e.g., Farmer & Vogel, 2013; He et al., 2011; Razavi et al., 2013). With an increase availability of various different data types and computing power over the last couple of decades it is now possible to re-visit the challenging issues using data-driven approaches such as Machine Learning (Dibike &

Solomatine, 2001; Solomatine & Ostfeld, 2008), which also belongs to the second category of hydrostatistical methods. Farmer and Vogel (2013) summarized the general procedure of the second category as a three-step process. Step 1: selection of one or multiple donor gauges based on some measures of hydrologic similarity. Step 2: estimation of streamflow statistics, such as the mean and standard deviation, at the target location. Step 3: transference of the streamflow time series from the donor gauge(s) to the target site (e.g., partially gauged/incomplete or ungauged).

The accuracy of inferred daily streamflow estimations based on Step 3 is conditioned on the accuracy of a proper selection of the donor gauge(s) in Step 1. This selection is typically based on an assessment of the hydrologic similarity between the target and the donor gauge(s) and whether a single or multiple donor gauges are used. A number of approaches have been used so far with different levels of complexity, data requirements and accuracies (e.g., Archfield & Vogel, 2010; Arsenault & Brissette, 2014; Farmer & Vogel, 2013; Halverson & Fleming, 2015; Mishra & Coulibaly, 2009; Smakhtin, 1999; Smakhtin et al., 1997; Zhang & Chiew, 2009). Examples of alternative approaches available in the literature include multivariate analysis along with clustering algorithms for rationalizing a hydrometric network (Burn & Goulter, 1991; Mishra & Coulibaly, 2009) by identifying redundant stations; and the use of entropy (from information theory) to assess the hydrologic similarity in the design of hydrometric networks so that the stations are as independent of each other as possible (Mishra & Coulibaly, 2009) which also reduces redundancies and thus maximizes the information content of individual sites.

The approach of selecting the nearest gauge as the donor is a convenient and widely used method due to its simplicity and minimum data requirements (e.g., Asquith et al., 2006; Emerson et al., 2005; Farmer & Vogel, 2013; Mohamoud & M., 2008). For example, Farmer & Vogel, (2013) adopted this simple distance-based method in the donor selection procedure (Step 1) in their study where a number of methods using different streamflow statistics in Step 2 and Step 3 were investigated and compared. Archfield and Vogel (2010) developed a procedure called *Map correlation method* that uses time series from several streamflow gauges in the study area to create a correlation map based on a kriging method and then uses that map to estimate the correlation between a given ungauged location and nearby gauges. They concluded that (1) the distance-based approach does not provide a consistent selection criterion; (2) the most correlated gauge is not always the closest one by distance; and (3) the accuracy based on the most correlated gauge outperforms the one based on the distance in most cases. The correlation-based approach is generally better than the distance-based approach because the streamflow data is more effectively used in the correlation-based approach and the marginal independence between any pair of gauges can be easily determined (Koller & Friedman, 2009). Here, the pair-wise correlation between two gauges is used to evaluate their marginal independence. That is, the two gauges are assumed to be independent of each other if their pair-wise correlation is below a given threshold. For example, Halverson & Fleming (2015) used complex network theory to compute a graph, representing the marginal independence assumptions between the streamflow time series, by setting a correlation threshold of 0.7 in identifying whether two gauges in question are independent or not.

Although using a single donor gauge to estimate streamflow time series has been a dominant approach (Farmer, 2016) regardless of whether it is based on distance or correlation. Smakhtin et al. (1997) and Smakhtin (1999) proposed to use more than one donor gauge from nearby gauges to improve the streamflow estimations for ungauged basins. Zhang and Chiew (2009), and Arsenault and Brissette (2014) also concluded that the estimation from multiple donor gauges is more accurate in general than that from a single donor gauge case. In these studies, multiple donor gauges were investigated based on methods such as the degree of similarity of flow regimes between the donor and destination gauges, spatial proximity, physical similarity, simple arithmetic mean, inverse distance weighting, combinations of some of these methods, or an assignment of a fixed number of donor gauges. The challenges of these approaches include: (1) how to measure the similarity; (2) how to systematically determine which gauges should be the donor gauges; and (3) how many donor gauges each individual target gauge should have.

In addition to Archfield and Vogel (2010), other previous work (e.g., Farmer, 2016; Skøien & Blöschl, 2007; Solow & Gorelick, 1986) also showed that geostatistical methods, such as kriging that uses multiple donor gauges, are an effective alternative. The kriging method is an spatial interpolation technique that estimates values at target locations as a linear weighted combination of the observations from different locations. The weights are assigned based on a variogram model which is usually fitted based on the variance between observations as a function of the distance between locations. The kriging method avoids problems in terms of selecting the number of donor gauges and the individual donor gauges since all of them are used in a linear combination fashion. The kriging method is useful in transferring information from gauged to ungauged locations (Villeneuve et al., 1979). However, the accuracy of its estimation depends on the density and quality of the measurements of the gauged sites. Virdee & Kottegoda (1984) noticed that a major problem with kriging is the lack of data with needed density.

For a commonly encountered situation in which the density of streamflow network is sparse, kriging is not a good candidate. From the aforementioned various methods other than kriging, it appears that the correlation-based single donor method (i.e., pair-wise marginal independence approach) is less subjective and provides more consistent results while the multi-donor methods lead to better results with subjective selection process on the donor gauges. Therefore, it is critical to develop a method that is less subjective in selecting a set of multi-donor gauges for each target location (i.e., Step 1). In this study, we present a novel method which draws on the strengths of existing methods but overcomes their weaknesees. More specifically, we present an approach that can explictly and effectively consider the correlation structure of the entire gauge network rather than the pair-wise correlation between any two gauges used in the existing methods. The pair-wise correlation approach is basically a local approach that does not reflect the dependence structure of the daily streamflow distribution, based on conditional independence conditions in graphical models, embodied by the underlying streamflow network. Since the conditional independencies among gauges in the streamflow network are typically not apparent in the correlation matrix but in its inverse matrix, i.e., the precision matrix (Koller & Friedman, 2009), the existing pair-wise correlation-based methods on multi-donor selection process are ineffective. In this study, we use the precision matrix to extract dependence structure of the gauge network based on the concept of conditional independence conditions in graphical models. We then use such identified dependence information to select donor gauges (Step 1). Since the donor gauges are selected based on the dependence structure of the entire gauge network, our method can be considered as a global approach as opposed to the existing local approach where only pair-wise correlations between gauges in the network are considered in which the marginal independence assumption is applied. Our method is generic, flexible and also more effective since it can extract implicit information (i.e., conditional independence structure of the underlying streamflow network) using a sparse precision matrix instead of the commonly used correlation matrix. With this new method, we can infer daily streamflow for active gauges with data gaps and extend data for inactive gauges which are defined as those that are no longer collecting data but collected data in the past (i.e., data extension). In addition, with this new method of filling in data gaps and extending data records, we can estimate daily streamflow at ungauged sites or improve the estimation of daily streamflow at ungauged sites based on the kriging method. Furthermore, a new algorithm based on the conditional independence concept is presented to remove gauges, when required, from an existing streamflow network with the least loss of information. That is, the closure of stations should be the ones where they can be estimated from other stations (Villeneuve et al., 1979).

The remainder of this paper is organized as follows. Section 2 briefly describes widely used approaches to transfer streamflow from a single donor to a target basin. Section 3 presents our new approach to systematically identify multiple donors based on the precision matrix and a new framework to infer daily streamflow time series based on the selected set of donor gauges. In addition, a new and general method to remove streamflow gauges from the existing hydrometric network with the least loss of information is presented in this section. Section 4 presents an example to illustrate and evaluate the new approaches presented in Section 3. Section 5 presents the results and discussions. Finally, Section 6 provides a summary of the main findings from this work.

## 2. Common approaches

This section briefly describes some common approaches used to transfer streamflow from a single donor to a target gauge (Step 2 and Step 3), assuming the single target gauge is already identified by a method from Step 1. Let $Q_j$ and $Q_i$ represent the streamflow from a target and a donor gauge, respectively, and assume that the estimated streamflow time series at the target location ($\widehat{Q}_j$) is obtained by transferring the streamflow time series from a single donor gauged catchment by a scaling function such that $\widehat{Q}_j$ is an approximation of $Q_j$ (e.g., Archfield & Vogel, 2010; Farmer & Vogel, 2013).

### 2.1 Drainage Area Ratio

The drainage area ratio (DAR) is a simple scaling procedure that only requires the areas from the target and donor catchments along with the streamflow time series from the donor gauge:

$$\widehat{Q}_j = \frac{A_j}{A_i} Q_i \qquad (1)$$

where $A_j$ and $A_i$ are the areas of the target and donor catchments, respectively. The DAR method represented by equation (1) assumes that the discharge per unit area is the same between the target $Q_j$ and donor $Q_i$ catchments at the same time step. This method is effective if the climate and hydrologic regimes at the target and donor sites are similar and the area is the only dominant factor affecting the streamflow. However, such requirements are generally not met, because a number of factors can significantly change the scaling relationship in Eq. (1), such as orographic effects where the site at a different elevation is likely to receive a different amount of rainfall and thus a different amount of runoff per unit area; a site in the windward side of the mountain versus the site in the rain shadow side where the rainfall characteristics are dramatically different; differences in slopes, soil types, land cover and land use which can affect the conditions of runoff generation, leading to differences in the basin's response to rainfall; and differences in temperature that affect the evapotranspiration losses and runoff per unit area.

## 2.2 Scaling by the Mean (SM)

Scaling by the mean (SM), also called *Standardization by the mean streamflow* (Farmer & Vogel, 2013), is a method that requires the mean streamflow from the target and donor gauges in addition to the streamflow time series from the donor gauge:

$$\widehat{Q}_j = \frac{\mu_j}{\mu_i} Q_i \qquad (2)$$

where $\mu_j$ and $\mu_i$ are the mean streamflow of the target and donor gauges, respectively. The SM method represented by equation (2) assumes that the discharge scaled by the mean streamflow is the same between the target $Q_j$ and donor $Q_i$ catchments at the same time step.

## 2.3 Scaling by the Mean and Standard deviation (SMS)

Scaling by the mean and standard deviation (SMS), also called *standardization with mean and standard deviation* (Farmer & Vogel, 2013), is a method that requires information of the mean and standard deviation from the streamflow for the target and donor gauges in addition to the streamflow time series from the donor catchment. This method was originally presented by Hirsch (1979) and termed as maintenance of variance extension (MOVE) and more recently reported by Archfield & Vogel, (2010), and Farmer & Vogel, (2013) as:

$$\widehat{Q}_j = \frac{\sigma_j}{\sigma_i}(Q_i - \mu_i) + \mu_j \qquad (3)$$

where $\mu_j, \mu_i, \sigma_j$ and $\sigma_i$ are the mean streamflow of the target and donor gauges and the standard deviation of the target and donor gauges, respectively. The SMS method represented by equation (3) assumes that the discharge scaled by the mean and standard deviation from the streamflow is the same between the target $Q_j$ and donor $Q_i$ catchments at the same time step.

## 2.4 Linear Regression

The linear regression method between streamflow time series of the target and donor gauges is rarely used as a transfer method of streamflow due to the lack of streamflow data for the target catchment. However, this information is available for the case of inactive gauges. The regression method (REG) is a simple least squares linear regression (e.g., Sachindra et al., 2013) where the regression coefficients for the slope $\gamma_{ij}$ and the intercept $\gamma_{0j}$ are estimated according to equation (4):

$$\widehat{Q}_j = \gamma_{0j} + \gamma_{ij} \cdot Q_i \tag{4}$$

## 3. New approach of selecting multiple donor gauges via graphical models

This section presents a novel approach that describes how our new algorithms (1) select a set of multiple donor gauges for each target location over a study area (Step 1) by using precision matrix obtained with a sparse graphical model; (2) estimate a matrix of regression coefficients that allow simultaneous inference of streamflow from the selected set of donor gauges to target gauges (Step 2); and (3) perform the inference of the daily streamflow time series (Step 3). The streamflow inference is based on minimizing two main objectives: (1) the streamflow estimation error and (2) the model complexity. Here, model complexity refers to the number of donor gauges required to infer the streamflow at a given target location. Thus, the simplest model would be to select a single donor gauge while the most complex model would be to select all of the available gauges in the network. We argue that there is a trade-off between the model complexity and the accuracy of the estimation, and that a complex model does not necessarily always result in more accurate estimated streamflow time series than a simpler model due to noises. This study aims at finding a suitable balance between the number of donor gauges and the accuracy through optimizing the two objectives -- error and complexity.

The remaining of this section is structured as follows. The building blocks for the development of our approach are described in sub-sections 3.1 to 3.4. Our algorithm which selects a sparse model with a low validation error (i.e., determination of model complexity), "Selection of Graph Model" (called SGM algorithm hereafter), is provided in sub-section 3.5. The model complexity determined in sub-section 3.5 is then used to train a multiple linear regression model. The inference of daily streamflow given the graph selected by the SGM algorithm is described in sub-section 3.6. Based on sub-sections 3.5 and 3.6, we develop a new algorithm, "Removal of Streamflow Gauges" (called RG algorithm hereafter), to remove gauges from the hydrometric network with the least loss of information.

**3.1 Multiple Linear Regression (MLR)**

A simple multiple linear regression (MLR) approach is to extend the single linear regression of equation (4). That is, for a set of *p* available gauges with daily streamflow records over the study area, each location, *j*, assumed as a target ungauged location, can be expressed by equation (5) as follows, while all the remaining gauges constitute the set of donor gauges:

$$\widehat{Q}_j = \eta_{0j} + \sum_{i=1, i \neq j}^{p} \eta_{ij} \cdot Q_i \tag{5}$$

where the estimated streamflow time series $\widehat{Q}_j$ at a target location, *j*, is computed by a linear combination of (*p* - 1) donor gauges, and $\eta_{ij}$ and $\eta_{0j}$ represent the multiple regression coefficients (slopes and intercept). Notice that because the donor and target gauges must be different, the target location, *j*, must be different from the donor location, *i*, and that all the available donor gauges are used.

Since the probability distribution of streamflow is often well approximated by a log-normal distribution (e.g., Stedinger, 1980), equation (5) can be modified and expressed by equation (6) in which $Y_i$ follows a normal distribution and is related to $Q_j$ by a logarithmic transformation, where $\rho_{ij}$ and $\rho_{0j}$ represent the multiple regression coefficients (slopes and intercept).

$$\widehat{Y}_j = \rho_{0j} + \sum_{i=1, i \neq j}^{p} \rho_{ij} \cdot Y_i \qquad (6)$$

To avoid numerical issues with the logarithm of zero-valued streamflow, Farmer (2016) assigned a small constant value (e.g. 0.00003 m^3/s), smaller than any non-zero value in the data set, to the zero-valued streamflow when applying a logarithmic transformation to the streamflow time series. Here, a different approach is followed. A value of one is added to the daily streamflow time series before the logarithmic transformation is performed, as expressed by equation (7), due to the fact that zero-valued streamflow is mapped to zero in the log-transformed variable and the transformation is reversible without loss of precision. Nevertheless, the results of applying either Farmer's approach or the approach of Eq. (7) are almost identical as shown in this study (see section 3.5.2).

$$Y_i = log(Q_i + 1) \qquad (7)$$

For convenience and simplicity, the standard score (Z-score) is used to define a new variable **Z**, as expressed by equation (8) where $\mu_{y_i}$ and $\sigma_{y_i}$ are the mean and standard deviation of $Y_i$.

Therefore, each vector $Z_i$ has a mean of zero and a standard deviation of one. Equation (9) represents a Z-score regression, where the intercept is zero and the regression coefficient $\alpha_{ij}$ is the correlation between the *jth* target $Z_j$ and the *ith* donor gauge $Z_i$ (z-score of log-transformed) streamflow time series.

$$Z_i = \frac{Y_i - \mu_{y_i}}{\sigma_{y_i}} \tag{8}$$

$$\hat{Z}_j = \sum_{i=1, i \neq j}^{p} Z_i \cdot \alpha_{ij} \tag{9}$$

Note that equation (9) is valid for each of the *p* selected gauges. That is, the column vector $\hat{Z}_j$ is computed for $1 \leq j \leq p$. Equations (10) and (11) express equation (9) in a matrix form.

$$\hat{Z} = Z \cdot A \tag{10}$$

$$A = \begin{bmatrix} 0 & \alpha_{12} & \cdots & \alpha_{1p} \\ \alpha_{21} & 0 & \cdots & \alpha_{2p} \\ \cdots & \cdots & 0 & \cdots \\ \alpha_{p1} & \cdots & \cdots & 0 \end{bmatrix} \tag{11}$$

The linear system defined by equation (10) assumes that the *jth* gauge is the target and that the remaining (*p* - 1) gauges are the donor gauges for each of the *p* gauges. Equation (11) shows the elements of a *p* by *p* matrix **A** used in equation (10). The elements of matrix **A** are the regression coefficients $\alpha_{ij}$ in equation (9) and the *jth* column represents the vector of regression coefficients required to estimate the column vector $\hat{z}_j$. Note that the diagonal elements of **A** are zero. That is, $for\ 1 \leq j \leq p:\ \alpha_{jj} = 0$. $\hat{Z}$ is a matrix with the estimated streamflows computed from the observed streamflow **Z** that follows a standard normal distribution, and the squared matrix of regression coefficients **A**. $\hat{Z}$ and **Z** are *n* by *p* matrices where *n* is the number of daily streamflow records. Note that if the streamflow data do not follow the log-normal distribution as assumed here, one can easily transform the data into the log-normal distribution, and thus equations (6)-(11) are applicable.

## 3.2 Concept of Gaussian Models and MLR

**Z** is a multivariate normal distribution over *p* random variables with covariance matrix $\Sigma$ and zero mean vector such that $\mathbf{Z} = (\mathbf{z_1}, \ldots, \mathbf{z_p})$, where the random variable $\mathbf{z_j}$ represents the Z-score of the logarithm of the streamflow data at the *j*th gauge, therefore it follows a normal distribution with zero mean and unitary standard deviation for $1 \leq j \leq p$. **Z** defines an undirected graphical model known as a *Gaussian graphical model* where the underlying graph **G** is defined by a set of vertices **v** and edges **e** such that the graph $\mathbf{G} = (\mathbf{v},\mathbf{e})$ represents (conditional) independence assumptions among the random variables. This conditional independence means that if there is not an edge on the graph **G** between the i*th* and j*th* location, then these two gauges are independent from each other given the remaining gauges. The existence of such conditional independence implies that for a given target location some of the donor gauges are redundant or not directly correlated to the target location and therefore, they can be removed from the set of donor gauges for the target location under consideration. Equation (10) is a general relationship between each possible target and donor gauges. However, it does not explicitly show how to compute the matrix of regression coefficients **A**. One simple way would be to use an approach from the previous subsection. That is, computing the elements of the matrix **A**, column by column, by means of MLR as shown in equation (9) for each of the *p* target locations. However, this approach assumes that all of the *(p - 1)* donor gauges are included in the regression for each target location. That is, it implies of having a graph **G** where each vertex is connected to all of the remaining vertices. In other words, it means a complete graph with $\frac{p^2-p}{2}$ edges. Therefore, this approach does not satisfy our second design objective of minimizing the model complexity by reducing the number of donor gauges, whch is equivalent to reducing the number of edges of the underlying graph **G**. Furthermore, there are two problems with dense graphs. First, a more complex graph requires more training data to avoid overfitting. Second, only a single gauge or a small number of gauges could be removed from the hydrometric network because each gauge is estimated based on all of the remaining gauges connected in the complex graph. Keeping a good balance between the complexity of a graph and the accuracy of an estimation is the goal of our new method. This is achieved by promoting sparsity while minimizing the estimation error.

### 3.3 Relationship between MLR, the covariance, and the precision matrices

This subsection describes two methods to compute the regression coefficients of matrix **A** represented by equation (11). The first method is based on the covariance matrix $\boldsymbol{\Sigma}$ and the second one is based on the inverse of the covariance matrix which is called the precision matrix $\boldsymbol{\Theta}$ as expressed by equation (12),

$$\boldsymbol{\Theta} = \boldsymbol{\Sigma}^{-1} \tag{12}$$

Since the true covariance ($\boldsymbol{\Sigma}$) or precision ($\boldsymbol{\Theta}$) matrices are unknown, **A** can only be estimated from the noisy *p*-dimensional observed data from **Z** which often follow a normal distribution. One method is based on an estimated covariance matrix, represented by **W**, and the other method is based on an estimated precision matrix, represented by $\widehat{\boldsymbol{\Theta}}$. Following Friedman et al. (2008), the columns and rows of **W** can be permuted so that the target *j*th gauge is the last and then partition the matrices into four blocks composed by a square submatrix $\mathbf{W_{11}}$ with (*p* - *1*) columns (and rows), a column vector $\mathbf{w_{12}}$ with (*p* - *1*) elements, its transposed (row) vector $\boldsymbol{w_{12}^T}$ and a scalar $w_{22}$. Similar partition scheme leads to the estimated precision matrix $\widehat{\Theta}$ to define the four blocks $\widehat{\Theta}_{11}, \widehat{\theta}_{12}, \widehat{\theta}_{12}^T$ and $\hat{\theta}_{22}$, respectively.

The relationship between the estimated covariance $\mathbf{W}$, the estimated precision $\widehat{\Theta}$ and the $p$ by $p$ identity matrix $\mathbf{I}$ is represented by $\mathbf{W} \cdot \widehat{\Theta} = \mathbf{I}$. The block-wise expansion of this equation, adapted from Friedman et al., (2008), leads to equation (13) as follows:

$$\begin{pmatrix} \mathbf{W}_{11} & \mathbf{w}_{12} \\ \mathbf{w}_{12}^T & w_{22} \end{pmatrix} \begin{pmatrix} \widehat{\Theta}_{11} & \widehat{\theta}_{12} \\ \widehat{\theta}_{12}^T & \hat{\theta}_{22} \end{pmatrix} = \begin{pmatrix} \mathbf{I} & \mathbf{0} \\ \mathbf{0}^T & 1 \end{pmatrix} \qquad (13)$$

Equation (14) shows the column decomposition of the matrix $\mathbf{A}$:

$$\mathbf{A} = [\boldsymbol{\alpha}_1 \quad \cdots \quad \boldsymbol{\alpha}_j \quad \cdots \quad \boldsymbol{\alpha}_p] \qquad (14)$$

There are several ways to compute the regression coefficients for each column of the matrix $\mathbf{A}$. Equation (15) shows a method based on the estimation of the covariance matrix $\mathbf{W}$ and the partitioned matrices from equation (13). It computes $\boldsymbol{\alpha}_j$ (equation (14)) for $1 \leq j \leq p$ using $\mathbf{W}_{11}$ as the predictor matrix and $\mathbf{w}_{12}$ as the response vector.

$$\boldsymbol{\alpha}_j = \mathbf{W}_{11}^{-1} \cdot \mathbf{w}_{12} \qquad (15)$$

Alternatively, the regression coefficients can be computed from the estimation of the precision matrix $\widehat{\Theta}$. Equation (16) shows the results with the latter approach.

$$\boldsymbol{\alpha}_j = -\frac{1}{\hat{\theta}_{22}} \widehat{\boldsymbol{\theta}}_{12} \qquad (16)$$

Equation (16) is derived by expanding the product from the first row and second column of Equation (13) such that $\mathbf{W}_{11} \cdot \widehat{\boldsymbol{\theta}}_{12} + \mathbf{w}_{12} \cdot \hat{\theta}_{22} = \mathbf{0}$. After some algebra manipulation, one can obtain equation (16). Equation (16) shows how the precision matrix $\widehat{\Theta}$ and the matrix of regression coefficients, $\mathbf{A}$, are related to each other. The elements of the matrix $\mathbf{A}$ are computed as $\boldsymbol{\alpha_j}$ for $1 \leq j \leq p$. The vector of regression coefficients $\boldsymbol{\alpha_j}$ for the *j*th column of **A** is proportional to the vector $\widehat{\boldsymbol{\theta}}_{12}$ of $\widehat{\boldsymbol{\Theta}}$. The matrix $\widehat{\boldsymbol{\Theta}}$ is closely related to the representation of the underlying graphical model **G** as zero elements in $\widehat{\boldsymbol{\Theta}}$ represent the missing edges in the graph **G**.

Thus, graph **G** can be represented by an adjacency matrix defined by equation (17) below where $g_{ij}$ and $\widehat{\theta}_{ij}$ represent the element of the *ith* row and *jth* column of **G** and $\widehat{\boldsymbol{\Theta}}$, respectively.

$$\mathbf{G} = \begin{cases} g_{ij} = 1 & if \quad |\hat{\theta}_{ij}| > 0 \\ g_{ij} = 0 & otherwise \end{cases} \quad (17)$$

Given that **Z** is a zero-mean vector, the calculation of the empirical covariance matrix **S** is simplified to equation (18) and the empirical precision matrix **T**, an inverse of matrix **S**, is defined by equation (19).

$$\mathbf{S} = \frac{1}{n-1} \mathbf{Z}^\mathbf{T} \cdot \mathbf{Z} \quad (18)$$

$$\mathbf{T} = \mathbf{S}^{-1} \quad (19)$$

Equation (20) shows how to calculate each column of the matrix **A** by replacing the estimated covariance matrix **W** in equation (15) with the empirical covariance matrix **S**, while equation (21) shows how to compute each column of matrix **A** by replacing $\widehat{\boldsymbol{\Theta}}$ with **T** in equation (16).

$$\boldsymbol{\alpha_j} = \mathbf{S_{11}}^{-1} \cdot \mathbf{s_{12}} \quad (20)$$

$$\boldsymbol{\alpha_j} = -\frac{1}{t_{22}} \mathbf{t_{12}} \quad (21)$$

Even though **T** is calculated by inverting **S**, Equation (21) is more efficient than equation (20) because it does not require the inversion of any additional matrix when **T** is known. In this work we describe a way to avoid computing the full empirical precision matrix **T**, but to compute a sparse precision matrix instead. Using equation (21) is perhaps the fastest way to estimate the coefficients of matrix **A** and, therefore, the inferred streamflow from equation (10). This approach represents the case where each gauge is inferred based on the ($p - 1$) remaining gauges through the MLR method. Thus, the apparent computational efficiency of the method is achieved at the expense of the model complexity. If **G** is sparse, however, then the conditional independence assumptions imply that the precision matrix should also be sparse. In practice, both the covariance matrix **Σ** and the precision matrix **Θ** are unknown and thus, they are approximated by the empirical covariance matrix **S** and the empirical precision matrix **T** based on a finite number of noisy observations. The empirical precision matrix **T** obtained is generally not sparse d'Aspremont et al., (2008) due to the nature of the noisy data. Hence, the underlying graph **G** from the *Gaussian graphical model* is not sparse but a complete graph where each gauge depends (conditionally) on all of the remaining gauges in the hydrometric network. The MLR approach is thus often times associated with a complex model as MLR tries to use all of the predictor variables from a complete graph **G**. Since the objective is to infer the streamflow values at the target location with limited errors by selecting certain gauges in the network as the donor gauges, it is appropriate to simply select the most relevant donor gauges to be included as the predictors. This is equivalent to making the graph **G** sparse. Therefore, our approach is to remove the least important edges from the graph **G** through a *Gaussian graphical model* by applying an algorithm known as the *Graphical Lasso*, through which we build a sparse graph while keeping a relatively low estimation error for the inferred streamflow values.

### 3.4 The Graphical Lasso

The *Graphical Lasso* (Glasso) is an algorithm defined initially by Friedman et al., (2008) which imposes sparsity to the precision matrix by tuning a parameter $\lambda$. This algorithm has been actively used, analyzed and improved by several authors (Mazumder & Hastie, 2012; Sojoudi, 2014; Witten et al., 2011). Our work used the glasso Matlab package (*glasso*) and also a more recent efficient implementation called *GLASSOFAST* (Sustik & Calderhead, 2012).

The *Glasso* algorithm implements an efficient solution to the problem by maximizing the Gaussian log-likelihood according to the formulation given in equation (22), adapted from Friedman et al., (2008), where *det* and *tr* are the determinant and trace of a square matrix respectively, $||\widehat{\Theta}||_1$ is the $L_1$ norm of estimated precision matrix $\widehat{\Theta}$ (i.e., the sum of the absolute value of all the elements in the matrix) and $\lambda$ is the $L_1$ norm regularization parameter.

$$\widehat{\Theta}_{Glasso} \equiv arg_{\widehat{\Theta}}max[log(det\ \widehat{\Theta}) - tr(\mathbf{S} \cdot \widehat{\Theta}) - \lambda||\widehat{\Theta}||_1] \tag{22}$$

The *Glasso* algorithm requires that the probability distribution of the input data be relatively well described by a multivariate Gaussian distribution as is the case for the multivariate random variable **Z**. The inputs required by the *Glasso* algorithm are the empirical covariance matrix **S** and the regularization parameter $\lambda$. The output from the *Glasso* algorithm is a potentially sparse precision matrix estimate $\widehat{\Theta}_{Glasso}$ optimized by equation (22). Equation (23) shows the inputs and output of the *Glasso* algorithm. The estimation of the regression coefficients of matrix **A** for the inference of streamflow time series via *Glasso* is achieved by applying equation (16) in which $\widehat{\Theta}$ is replaced by $\widehat{\Theta}_{Glasso}$ as shown in equation (24) below.

$$\widehat{\Theta}_{Glasso} = Glasso(\mathbf{S}, \lambda) \tag{23}$$

$$\boldsymbol{\alpha_j} = -\frac{1}{\widehat{\theta}_{Glasso_{22}}}\widehat{\boldsymbol{\theta}}_{Glasso_{12}} \tag{24}$$

If the regularization parameter $\lambda$ is equal to zero, the estimated precision matrix $\widehat{\Theta}_{Glasso}$ is equivalent to the empirical precision matrix **T** obtained by the (non-regularized) MLR approach with equation (19) and the corresponding graph **G** is a complete graph. On the other hand, if the regularization parameter is very large, the underlying graph **G** would have zero edges. An algorithm (SGM) is presented in subsection 3.5.6 to select the $\lambda$ parameter based on a multi-objective optimization procedure that minimizes the error metric and also the number of edges of the underlying sparse Gaussian Graphical Model.

## 3.5 Graphical Model Selection

Our approach in selecting a proper subset of donor gauges to be used for inferring each streamflow gauge (Step 1) is to apply the conditional independence assumptions encoded in the precision matrix. In other words, the idea of conditional independence is used to find a subset of donor gauges for each target location. This proposed approach promotes sparsity on the precision matrix and, therefore, leads to an underlying graph **G** with fewer edges which is consistent with the parsimonious principle. That is, a simpler model that explains well the observations should be preferred over more complex models. Under such a context, the parsimonious principle implies a selection of an underlying graphical model that is as sparse as possible while keeping the estimation error relatively low.

### 3.5.1 Imposition of sparsity to underlying graphical model

The sparsity is achieved by adjusting the regularization parameter $\lambda$ for the *Glasso* algorithm in conjunction with a thresholding procedure that uses an additional parameter $\tau$ defined by equation (25) below, which is a modification of equation (17).

$$\mathbf{G} = \begin{cases} g_{ij} = 1 & if \quad |\hat{\theta}_{ij}| > \tau \\ g_{ij} = 0 & otherwise \end{cases} \quad (25)$$

The thresholding procedure is required in addition to the $L_1$ norm regularization because even though the $L_1$ norm of the precision matrix decreases monotonically as $\lambda$ increases, the number of edges in the graph **G** does not necessarily decrease monotonically. Therefore, a multi-objective optimization is needed to minimize the mean error between the observed random variable **Z** and the inferred data matrix $\hat{\mathbf{Z}}$ from equation (10), and the number of edges of the underlying graph **G**. In addition to equation (25) for sparsity, there exist some situations where a particular edge from the *ith* to the *jth* gauge needs to be removed from the underlying graphical model by setting the element $g_{ij}$ to zero. One example of such situation is when both the *ith* and the *jth* gauges are known to be donor basins, therefore none of them need to be inferred and the corresponding edge in the graphical model should be removed. A similar case applies when both gauges are known to be the target gauges, the edge between them should not exist, as one gauge cannot be infered using the other as a donor. In such cases, the Glasso procedure with an optional parameter, graph **G**, in equation 26 allows removal of some edges. If that graph **G** is ommitted, as in Equation 23, it assumes that all edges are available. Therefore Equation 23 is equivalent to Equation 26, if graph **G** is a full graph. Equation 26 is also useful because it allows one to compute the sparse precision matrix with a prescribed sparsity pattern. In addition, if the regularization parameter $\lambda$ is equal to zero, then this equation is equivalent to a MLR where each target gauge is estimated by the donor gauges that share an edge with it in the graph **G**.

$$\widehat{\mathbf{\Theta}}_{train(\lambda,G)} = Glasso(\mathbf{S_{train}}, \lambda, \mathbf{G}) \tag{26}$$

3.5.2 Preparation of data sets

The normalized standard Gaussian (Z-score of log-transformed) daily streamflow data set, **Z**, is sorted in ascending order by the timestamp of each daily record and then divided into three disjoint sets of approximately same size. The subsets are used, respectively, for training $\mathbf{Z}_{train}$, validation $\mathbf{Z}_{val}$, and testing $\mathbf{Z}_{test}$. $\mathbf{Z}_{train}$ is used for training the inference model by computing the regression coefficients for matrix **A**. $\mathbf{Z}_{val}$ is used for choosing the $\lambda$ and $\tau$ values that minimize the validation error and the number of edges of the underlying graph **G,** and $\mathbf{Z}_{test}$ is used for assessing the predictive capability of the streamflow inference algorithm through estimating the error based on the new data. The least recent two thirds of the daily streamflow records are randomly assigned to the training $\mathbf{Z}_{train}$ and validation $\mathbf{Z}_{val}$ data sets with a split ratio of 50%. The remaining one third of the data (most recent) is used as the test set $\mathbf{Z}_{test}$.

3.5.3 Estimation of training covariance and sparse precision matrices

The initial training precision matrix, $\widehat{\Theta}_{train(\lambda, G_{full})}$, for a given value of the regularization parameter $\lambda$, is computed by applying the *Glasso* algorithm of equation (23) using $\mathbf{S}_{train}$. The training covariance matrix, $\mathbf{S}_{train}$, was estimated by applying equation (18) along with the training dataset $\mathbf{Z}_{train}$. Alternatively, the initial precision matrix can be computed by using equation (26) with **G** equals to the full graph, $G_{full}$. The initial sparsity of the training precision matrix, $\widehat{\Theta}_{train(\lambda, G_{full})}$, is determined by the regularization parameter $\lambda$. Additional sparsity is achieved by computing a sparse graph, $G_\tau$, where a thresholding procedure for a given value of the truncation parameter $\tau$, as defined in equation (25), is applied using the precision matrix, $\widehat{\Theta}_{train(\lambda, G_{full})}$, obtained from the previous step. A new training precision matrix $\widehat{\Theta}_{train(\lambda, G_\tau)}$ is then computed using equation (26) and the sparse graph $G_\tau$. This sparse precision matrix has a value of zero on all elements where the graph $G_\tau$ has missing edges.

3.5.4 Estimation of regression coefficients and streamflow validation

The training matrix of regression coefficients, $\mathbf{A}_{train}$, is computed by the matrix decomposition of the training sparse precision matrix, $\widehat{\Theta}_{train(\lambda, G_\tau)}$, using equation (24), for $1 \leq j \leq p$, where $j$ is the *jth* gauge.

The standardized validation (Z-score of log-transformed) streamflow time series, $\hat{\mathbf{Z}}_{val}$, are estimated by using $\mathbf{A}_{train}$ and the validation dataset, $\mathbf{Z}_{val}$, as expressed in equation (27) below:

$$\hat{\mathbf{Z}}_{val} = \mathbf{Z}_{val} \cdot \mathbf{A}_{train} \tag{27}$$

The estimated log-transformed validation streamflow data, $\hat{\mathbf{Y}}_{val}$, is calculated using $\hat{\mathbf{Z}}_{val}$ in equation (8) and is shown in equation (28) below, for $1 \leq j \leq p$, where $j$ is the $jth$ gauge, $\mu_{y_{val_j}}$ and $\sigma_{y_{val_j}}$, represents, respectively, the mean and standard deviation of the vector $\hat{\mathbf{Y}}_{val_j}$.

$$\hat{\mathbf{Y}}_{val_j} = \hat{\mathbf{Z}}_{val_j} \cdot \sigma_{y_{val_j}} + \mu_{y_{val_j}} \tag{28}$$

The estimated validation streamflow data, $\hat{\mathbf{Q}}_{val}$, is calculated by applying the exponential function to $\hat{\mathbf{Y}}_{val_j}$, as shown in equation (29), for $1 \leq j \leq p$, where $j$ is the $jth$ gauge:

$$\hat{\mathbf{Q}}_{val_j} = exp\left(\hat{\mathbf{Y}}_{val_j}\right) - 1 \tag{29}$$

3.5.5 Score function and validation error

Selection of the graphical model should maximize the quality of the inferred daily streamflow time series. The goal is to estimate daily streamflow time series at the target gauges as accurate as possible so that these gauges can be potentially removed from the hydrometric network with the least loss of information. The score function is designed to measure the accuracy of the inferred values at the target sites. Equation (30) defines a conditional goodness-of-fit metric that calculates the value of the coefficient of determination $R^2$ between the observed and estimated $jth$ daily streamflow time series for the validation data set, where $R^2_{val_j}$ is the coefficient of determination, i.e., the square value of the correlation coefficient $R^2$, between the observed streamflow $\mathbf{Q}_{val_j}$ used for validation and the estimated streamflow $\hat{\mathbf{Q}}_{val_j}$, for $1 \leq j \leq q$, where $j$ is an index representing the *jth* gauge and *q* is the number of inferred gauges. By default, all of the gauges are considered as potential target sites, where *q* is equal to *p*. The score is positive if $R^2_{val_j}$ is greater than an assigned threshold $\Gamma$, otherwise, it is taken as zero. In this work the value of the threshold $\Gamma$ was set to 0.7. Equation (31) calculates the validation score. Equation (32) defines the validation error used in our multi-objective optimization procedure.

$$score_{val_j} = \begin{cases} R^2_{val_j} = R^2\left(\mathbf{Q}_{val_j}, \widehat{\mathbf{Q}}_{val_j}\right) & if \quad R^2_{val_j} > \Gamma \\ 0 & otherwise \end{cases} \quad (30)$$

$$score_{val} = \sum_{j=1}^{q} score_{val_j}, \quad q \leq p \quad (31)$$

$$error_{val} = \frac{q - score_{val}}{q} \quad (32)$$

While the number of edges of the underlying graph indicates its sparseness, the validation error of the graphical model is selected in such a way that it will maximize the validation score. The value of this validation error ranges over [0, 1] and is scale independent. It decreases as the validation score increases.

3.5.6 Selection of Graph Model Algorithm (SGM)

An algorithm called *Selection of Graph Model* (SGM) is developed to obtain an optimal underlying graph. A graph determined by the SGM algorithm is represented by $\boldsymbol{G_{sgm}}$. The SGM algorithm implements a multi-objective optimization procedure where the optimization objectives include: (1) minimizing validation error calculated by equation (32), and (2) minimizing the number of edges of the underlying graph. SGM generates a set of values for the regularization parameter $\lambda$ between a minimum value of $\lambda_{min}$ and a maximum value of $\lambda_{max}$.

For each regularization parameter value of $\lambda$, the truncation parameter, $\tau$, in Equation (25) is selected in such a way that the underlying graph has a given number of edges between a minimum, $K_{min}$, and a maximum, $K_{max}$, respectively. Given the multi-objective nature of the problem, a set of graphs corresponding to a set of non-dominated solutions on the Pareto front instead of a single solution is selected. Graph $\boldsymbol{G_{sgm}}$ thus represents one of the graphs from the set. A final graph(s) $\boldsymbol{G_{sgm}}$ is (are) selected from the set of candidate solutions as the one (ones) that offers (offer) desired trade-offs between the error and model complexity.

*Algorithm 1* below briefly describes a code implementation of the SGM algorithm. The parameter *res* is an integer number that represents the resolution of a sequence of sampling values to create a (1 x *res*) vector ***lamba_set*** with values between $\lambda_{min}$ and $\lambda_{max}$. *DonorSet* and *TargetSet* are optional parameters that represent a set of identifiers of the gauges that are known to be donors or targets, respectively. The default values for *DonorSet* and *TargetSet, in Algorithm 1,* are empty sets. That is, any gauge can potentially be used as a *Donor* or *Target* gauge. If *DonorSet* or *TargetSet* are non-null sets, then the corresponding gauges are treated as donor gauges or target gauges, respectively. Therefore, computing the graph model $\mathbf{G}_\tau$ defined in equation (25) implies removing all the edges between the *ith* and *jth* gauge when both, *i* and *j,* belong to *DonorSet* or both belong to *TargetSet.* The *getSequence* function generates the vector ***lamba_set***. A simple way to implement this function is by using a linear sequence. This algorithm is summarized below as Algorithm 1.

**Algorithm 1:** *Selection of Graph Model (SGM)*

**STEP 0.** Define the SGM inputs (assignment of default values)
    $\lambda_{min} = 0.01$; $\lambda_{max} = 0.10$; $K_{min} = 10$; $K_{max} = \frac{p^2 - p}{2}$; *res = 30;* $\Gamma = 0.7$
    *DonorGroup:={}; TargetGroup:={}*
    Retrieve training ($\mathbf{Z}_{train}$) and validation ($\mathbf{Z}_{val}$) data sets;

**STEP 1**. Compute the empirical covariance matrix using equation (18) from the training set:
$$n_{train} = length(\mathbf{S}_{train})$$
$$\mathbf{S}_{train} = \frac{1}{n_{train}-1}\mathbf{Z}_{train}^{\mathbf{T}} \cdot \mathbf{Z}_{train};$$

**STEP 2.** Generate Multi-objective optimization sampling points:
  *lambda_set* = *getSequence*(*minVal*=$\lambda_{min}$, *maxVal* = $\lambda_{max}$, *res*);
  **for** *r=1* **to** *res*:
    $\lambda_r$= *lambda_set[r]*;
    Compute the initial precision matrix from $\mathbf{S}_{train}$ using equation (23):
    $\widehat{\mathbf{\Theta}}_{train_r} = Glasso(\mathbf{S}_{train}, \lambda_r);$
    **for** $k= K_{min}$ **to** $K_{max}$:
      choose $\tau_{r,k}$ to compute the underlying graph model with at most $k$ edges using equation (25):
      $$\mathbf{G}_{r,k} = \begin{cases} g_{r,k_{ij}} = 1 & if \quad |\widehat{\theta}_{train_{r_{ij}}}| > \tau_{r,k} \\ g_{r,k_{ij}} = 0 & otherwise \end{cases};$$
      Compute the sparse training precision matrix, using equation (26):
      $\widehat{\mathbf{\Theta}}_{train_{r,k}} = Glasso(\mathbf{S_{train}}, \lambda_r, \mathbf{G}_{r,k});$
      Compute the training matrix of regression coefficients $\mathbf{A}_{train_{r,k}}$ from $\widehat{\mathbf{\Theta}}_{train_{r,k}}$, using equation (24), for $1 \leq j \leq p$:
      $$\boldsymbol{\alpha}_{train_{r,k_j}} = -\frac{1}{\widehat{\theta}_{train_{r,k_{22}}}}\widehat{\boldsymbol{\theta}}_{train_{r,k_{12}}};$$
      Compute the inferred Z-score log-transformed validation streamflow from equation (27):
      $\widehat{\mathbf{Z}}_{val_{r,k}} = \mathbf{Z}_{val} \cdot \mathbf{A}_{train_{r,k}};$
      Compute the inferred log-transformed validation streamflow using equation (28) for $1 \leq j \leq p$:
      $\widehat{\mathbf{Y}}_{val_{r,k_j}} = \widehat{\mathbf{Z}}_{val_{r,k_j}} \cdot \sigma_{y_{val_j}} + \mu_{y_{val_j}};$
      Compute the inferred validation streamflow using equation (29) for $1 \leq j \leq p$:
      $\widehat{\mathbf{Q}}_{val_{r,k_j}} = exp\left(\widehat{\mathbf{Y}}_{val_{r,k_j}}\right) - 1;$
      Calculate the validation score using equation (30) and equation (31) for $1 \leq j \leq q$:
      $$score_{val_{r,k_j}} = \begin{cases} R^2_{val_j} = R^2\left(\mathbf{Q}_{val_j}, \widehat{\mathbf{Q}}_{val_j}\right) & if \quad R^2_{val_j} > \Gamma \\ 0 & otherwise \end{cases}$$
      $score_{val_{r,k}} = \sum_{j=1}^{q} score_{val_{r,k_j}}, \quad q \leq p;$
      Calculate the validation error using equation (32):
      $error_{val_{r,k}} = \frac{q - score_{val_{r,k}}}{q};$
    store the sampling results: *multi_objective_points* = [$k$, $error_{val_{r,k}}$], $\lambda_r$ and $\mathbf{G}_{r,k}$.

**STEP 3.** Select the set of non-dominated solutions from *multi_objective_points*

**STEP 4.** From the set of non-dominated solutions, select a sparse graph (as the output), $\boldsymbol{G}_{sgm}$, with a suitable tradeoff between the number of edges and validation error and optionally the corresponding matrix of regression coefficients $\boldsymbol{A}_{sgm}$.

## 3.6 Stream flow inference

The inference task is greatly simplified once the underlying graph $G_{sgm}$ is identified by the SGM algorithm. This graph $G_{sgm}$ reveals conditional independent conditions between the streamflow gauges for the given hydrometric streamflow network. Therefore, a set of donor gauges best for each streamflow gauge is explicitly indicated by the graph $G_{sgm}$. Such a set includes only the donor gauges for which each target station conditionally depends on.

3.6.1 Inference of daily streamflow time series with graph $G_{sgm}$

Let matrix $A_{sgm}$ represent matrix **A** of Equation (11) whose element $\alpha_{ij}$ (i.e., regression coefficient) is determined based on graph $G_{sgm}$. The Z-score of the log-transformed streamflow time series for the test set $\hat{Z}_{test}$ can then be estimated directly using matrix $A_{sgm}$ and the test dataset $Z_{test}$. Thus, Equation (10) can be expressed as follows:

$$\hat{Z}_{test} = Z_{test} \cdot A_{sgm} \qquad (33)$$

To obtain $\hat{Y}_{test}$ from $\hat{Z}_{test}$, the mean $\mu_{Y_{test_j}}$ and standard deviation $\sigma_{Y_{test_j}}$ for the test set are required, but they are unknown. One way to overcome this problem is to assume that the mean and standard deviation for the test set are the same as those for the training set. Then, one can obtain $\hat{Y}_{test}$ from which to obtain the original streamflow time series $\hat{Q}_{test}$ by applying the exponential-transform function. Clearly, the assumption made here is not usually held.

An alternative approach is to perform an ordinary least squares multiple linear regression to estimate a new set of regression coefficients of $\beta_{ij}$ (slope) and $\beta_{0j}$ (intercept) over the log-transformed streamflow time series for the training data set over $1 \leq j \leq p$, using only the donors for the *jth* target site as expressed by Equation (34), where $donors(j) = donors(G_{sgm}, j)$.

$$\widehat{\mathbf{Y}}_{\text{test}_j} = \beta_{0j} + \sum_{i=1}^{size(donors(j))} \beta_{ij} \cdot \mathbf{Y}_{train_{donors(j)_i}} \qquad (34)$$

The daily streamflow time series, $\widehat{\mathbf{Q}}_{\text{test}_j}$, is estimated based on the log-transformed streamflow, $\widehat{\mathbf{Y}}_{\text{test}_j}$, for $1 \leq j \leq p$, as shown in equation (35).

$$\widehat{\mathbf{Q}}_{\text{test}_j} = exp\left(\widehat{\mathbf{Y}}_{\text{test}_j}\right) - 1 \qquad (35)$$

The third alternative is to directly apply MLR to the non-transformed streamflow time series avoiding the logarithmic transformation. Among these three approaches, results from Equation (34) should be either more accurate or more stable as indicated by Farmer (2016) who found that the logarithmic transformation of the streamflow is generally a more stable predictand than the streamflow itself.

3.6.2 Inference of daily streamflow time series using distance and correlation approaches

To evaluate the performance of our new method based on graph $\boldsymbol{G}_{sgm}$ in inferring daily streamflow time series, we compare our new method with two widely used methods, the distance-based method ("Dist") and the pair-wise correlation-based method ("Corr"). Two graphs, $\boldsymbol{G}_{dist}$ (distance-based) and $\boldsymbol{G}_{corr}$ (pair-wise correlation-based), are constructed. The $\boldsymbol{G}_{dist}$ graph is built starting with an empty graph (i.e., none of the gauges existed in the study region are connected) and then adding edges (i.e., connecting gauges) between each target site and its nearest neighbor site. In this case, each target site has one donor site, expressed as $\boldsymbol{G}_{dist,1}$, and the constructed graph structure is determined by the number of edges added and their relative locations in the gauge network. For the case of having two donor sites, edges between each target site and its nearest and second nearest neighbor sites are added in the graph, and is expressed as $G_{dist,2}$. Graph $G_{dist,3}$ represents the case where each target gauge has 3 donor sites. The graph of $G_{corr}$ is built in a similar way to $G_{dist}$ except that the most correlated sites instead of the nearest sites are selected. For the case with one donor site, the built graph is represented by $G_{corr,1}$. For the cases with two and three donor sites, the constructed graphs are represented by $G_{corr,2}$ and $G_{corr,3}$, respectively. The graphs of $G_{dist,i}$ and $G_{corr,i}$ ($i = 1, 2,$ and $3$) are built in such a way to mimic the current practice in which a fixed and equal number of donors for each target site is used in both distance- and correlation-based approaches. In comparison, an uneven number of donors for each target site is automatically determined and used in our new method. For both the distance- and correlation-based methods, the daily streamflow time series are inferred following the same procedure described in sub-section 3.6.1 as for our new method (i.e., SGM). The only difference is to replace the graph $G_{sgm}$ by $G_{dist,i}$ or $G_{corr,i}$ ($i = 1, 2,$ and $3$) in each case.

3.6.3 Estimation of test error

The test error is computed in the same way as the validation error described in sub-section 3.5.5, but using the test set as follows:

$$score_{test_j} = \begin{cases} R^2_{test_j} = R^2\left(\mathbf{Q}_{test_j}, \widehat{\mathbf{Q}}_{test_j}\right) & if \quad R^2_{test_j} > \Gamma \\ 0 & otherwise \end{cases} \quad (36)$$

$$score_{test} = \sum_{j=1}^{q} score_{test_j}, \quad q \leq p \quad (37)$$

$$error_{test} = \frac{q - score_{test}}{q} \quad (38)$$

3.6.4 Estimation of inference accuracy

The accuracy of each of the inferred gauges associated with the graphs from the SGM algorithm and with the graphs of $\boldsymbol{G}_{dist,i}$ and $\boldsymbol{G}_{corr,i}$ ($i = 1, 2,$ and $3$) is evaluated using the Nash–Sutcliffe efficiency coefficient (NSE) (Nash & Sutcliffe, 1970) with the testing data set. The NSE of the testing data set ($NSE_{test_j}$) is computed between the observed ($\mathbf{Q}_{test_j}$) and the inferred ($\widehat{\mathbf{Q}}_{test_j}$) streamflow time series for $1 \leq j \leq p$ as shown in equation (39) below.

$$NSE_{test_j} = NSE\left(\mathbf{Q}_{test_j}, \widehat{\mathbf{Q}}_{test_j}\right) \tag{39}$$

**3.7. Removal of streamflow gauges with the least loss of information**

The removal of streamflow gauges (RG) is a straightforward procedure once the model selection and inference stages are completed. The RG algorithm is designed to remove gauges that can be inferred by other gauges with the highest efficiency, i.e., with the highest NSE ($NSE_{test_j}$) for the testing data set. Thus, RG removes a gauge in the network with the highest $NSE_{test_j}$ first, and then marks the removed gauge as a "target gauge" and each of its neighbors as a "donor gauge". This process is repeated for the remaining available gauges in the network until all gauges are checked, with the exception of isolated gauges that should not be removed. Algorithm 2 below shows the details of the gauge removal process with the least loss of information.

Equation (40) and Equation (41) define a new score for the graph, based on the NSE, but it only includes the gauges that can be removed from the hydrometric network with the least loss of information according to our RG algorithm, and with an NSE value higher than the threshold $\boldsymbol{\Gamma}$. For this study the value for $\boldsymbol{\Gamma}$ was set to 0.7. The constant maxRemRank represents the maximum number of gauges removable from the RG algorithm for a given graph.

| Algorithm 2: Removal of Gauges (RG) Algorithm |
|---|
| **STEP 0.** Define RG inputs: $[NSE_{test_1}, ..., NSE_{test_q}]$, $\boldsymbol{G_{sgm}}$ |
| **STEP 1.** Initialize the rank of removal: rank=0 |
| **STEP 2.** Mark all the gauges with at least one edge as available for removal. Isolated nodes are marked as not available for removal. |
| **STEP 3.** Update the rank of removal (rank = rank + 1) |
| **STEP 4.** Define the *r*th gauge as the one with the highest Nash–Sutcliffe model efficiency coefficient from the currently available gauge set. Assign the *r*th gauge to the current rank of removal. |
| **STEP 5.** Mark the *r*th gauge and its neighbors on the underlying graph $\boldsymbol{G_{sgm}}$ as unavailable for removal. |
| **STEP 6.** Repeat from step 3 until there are no more available gauges for removal. |

The $\boldsymbol{graph\_score_{test}}$ is useful to assess the quality and quantity of the inference of daily streamflow time series for the removable gauges from a given graph model. The higher the graph_score$_{test}$ is, the better.

$$graph\_score_{test_{remRank}} = \begin{cases} NSE_{test_{remRank}} & if \quad NSE_{test_{remRank}} > \Gamma \\ 0 & otherwise \end{cases} \quad (40)$$

$$graph\_score_{test} = \sum_{remRank=1}^{maxRemRank} graph\_score_{test_{remRank}} \quad (41)$$

## 4. Study area and data sets

Our new method is applied to the Ohio River basin due to its size, relevance and good quality of long-term historical daily streamflow data. The Ohio River is the third largest river in terms of discharge in the United States. It is the largest tributary of the Mississippi River and accounts for more than 40% of the discharge of the Mississippi River (Benke & Cushing, 2011). The Ohio River is located between the 77° and 89° west longitude and between the 34° and 41° north latitude.

Table 1 lists the National Weather Service Location Identifier (NWSLI) which is used in this study to index each gauge, the drainage area of the corresponding sub-basin, and the USGS station identifier of the 34 streamflow gauges. The naturalized daily streamflow data are taken from the United States Geological Survey's (USGS) National Water Information System (NWIS: National Water Information System). This data set spans from January $1^{st}$, 1951 to December $31^{st}$, 1980 with a total of 10958 consecutive days (30 years) for all of the 34 streamflow gauges. There are no missing streamflow records for any day or gauge over the selected study period.

Table 1– List of 34 streamflow gauges over the Ohio River basin

| # | NWSLI | USGS STAID | Drainage Area (Km²) | # | NWSLI | USGS STAID | Drainage Area (Km²) |
|---|---|---|---|---|---|---|---|
| 1 | ALDW2 | 03183500 | 3,533 | 18 | GRYV2 | 03170000 | 777 |
| 2 | ALPI3 | 03275000 | 1,352 | 19 | KINT1 | 03434500 | 1,764 |
| 3 | ATHO1 | 03159500 | 2,442 | 20 | MROI3 | 03326500 | 1,766 |
| 4 | BAKI3 | 03364000 | 4,421 | 21 | NHSO1 | 03118500 | 453 |
| 5 | BELW2 | 03051000 | 1,052 | 22 | NWBI3 | 03360500 | 12,142 |
| 6 | BOOK2 | 03281500 | 1,870 | 23 | PRGO1 | 03219500 | 1,469 |
| 7 | BSNK2 | 03301500 | 3,364 | 24 | PSNW2 | 03069500 | 1,870 |
| 8 | BUCW2 | 03182500 | 1,399 | 25 | SERI3 | 03365500 | 6,063 |
| 9 | CLAI2 | 03379500 | 2,929 | 26 | SLMN6 | 03011020 | 4,165 |
| 10 | CLBK2 | 03307000 | 487 | 27 | SNCP1 | 03032500 | 1,368 |
| 11 | CRWI3 | 03339500 | 1,318 | 28 | STMI2 | 03345500 | 3,926 |
| 12 | CYCK2 | 03283500 | 938 | 29 | STRO1 | 04185000 | 1,062 |
| 13 | CYNK2 | 03252500 | 1,608 | 30 | UPPO1 | 04196500 | 772 |
| 14 | DBVO1 | 03230500 | 1,383 | 31 | VERO1 | 04199500 | 679 |
| 15 | ELRP1 | 03010500 | 1,424 | 32 | WTVO1 | 04193500 | 16,395 |
| 16 | FDYO1 | 04189000 | 896 | 33 | WUNO1 | 03237500 | 1,002 |
| 17 | GAXV2 | 03164000 | 2,929 | 34 | WYNI2 | 03380500 | 1,202 |

Following the procedure described in sub-section 3.5.2, the dataset was separated into 3 subsets. Data between 1951 and 1970 were used for "training" and "validation". The training data set consists of 50% of the data randomly selected from 1951 and 1970. The remaining data over the period of 1951 and 1970 consists of the validation set. The data between 1961 and 1970 was used as the "test" set.

## 5. Results and Discussion

### 5.1 Inference on streamflow

The inferred daily streamflow time series based on the new method (i.e., graph $G_{sgm}$) and the distance- and correlation-based methods (i.e., graphs of $G_{dist,i}$ and $G_{corr,i}$ with $i = 1, 2$, and $3$) are compared. For the latter two approaches, the three commonly used scenarios with 1, 2, and 3 donors per target gauge are considered. For our new method, the SGM algorithm was run with default parameters defined in *Algorithm 1*. That is, 30 different values of the regularization parameter $\lambda$ were used for graphs with edges between 10 (very sparse) and 561 (complete graph). Thus, the number of sampling points is (561 - (10-1)) *30 = 16560 (based on Step 2 of Algorithm 1).

The SGM algorithm selected 74 out of 16560 (0.45%) distinct graphs with different number of edges as the candidate solutions according to the multi-objective optimization procedure that minimizes both of the validation error and the number of edges. Figure 1 (a) shows results with trade-offs between the number of edges and the validation error, $error_{val}$, defined by Equation (32). The black dots represent the dominated solutions in the multiple-optimization space. The three red dots of the non-dominated solutions represent the graphs of SMG(25), SGM(47) and SGM(65) with 25, 47 and 65 edges, respectively. The remaining non-dominated solutions (i.e., solutions along the Pareto front) are represented by the green dots. Figure 1 (b) shows the comparison of the test error (Equation (38)) associated with the graphs, $G_{sgm}$, of SMG(25), SGM(47) and SGM(65), and graphs of $G_{dist,i}$ (distance-based) and $G_{corr,i}$ (correlation-based) with $i = 1, 2$, and $3$. More specifically, for the distance-based case, $G_{dist,1}$ = Dist(24), $G_{dist,2}$ = Dist(43), and $G_{dist,3}$ = Dist(65) with 24, 43, and 65 edges in each corresponding graph. For the correlation-based case, $G_{corr,1}$ = Corr(24), $G_{corr,2}$ = Corr(47), and $G_{corr,3}$ = Corr(68) with 24, 47, and 68 edges in each correspond graph as well.

At the top portion of the pareto front (e.g., green and red dots) in Figure 1 (a) a large validation error is present when the graphs are very sparse. But the error decreases quickly as the number of edges increases until about 44 edges from which point to about 92 edges, the change in error is negligible. At 93 edges there is a noticeable decrease in the validation error. From 93 to about 211 edges the change in validation error is negligible again. The next set of non-dominated solutions is from 211 edges onward with a slight decrease in the validation error where the pareto front becomes almost flat and reaches the minimum validation error at 222 edges. For this study region, it appears that a good trade-off between the sparsity and validation error is about having 44 or 45 edges, where the error is almost as low as the graph with 93 edges. Also, the error decreases dramatically at the beginning where an addition of a few more edges can significantly reduce the error. But for a graph with its number of edges starting around 45, an increase in the number of edges only reduces the error by a little bit. When the number of edges increases to about 93 or more, the improvement in error reduction becomes almost unnoticeable. Figure 1(a) shows that the relationship between the error and the number of edges has a L-like-shape in which the error approaches almost a constant when the graph reaches an edge number around 93. The few "sudden" discontinuities in Figure 1(a) are due to the nature of the error function which includes conditional terms above/below a threshold that might affect the total validation error once the threshold has been reached. The full graph with 561 edges is not in the set of non-dominated solutions, which means that using all of the gauges available in the network to infer the streamflow for the target site gives worse results than many of the sparser graphs. This is likely related to the noisy correlation calculated due to the large noises involved in the data. In fact, Figure 1 (a) shows that using graphs with more than 222 edges is unlikely to reduce the validation error anymore. This result clearly shows that it is not the more complex the better.

The three graphs SMG(25), SGM(47) and SGM(65), represented by the three red points in Figure 1 (a), were selected from a set of non-dominated solutions that, in terms of the number of edges, approximately matching the three graphs associated with 1-, 2-, and 3-nearest donors, Dist(24), Dist(43) and Dist(65) and the three graphs associated with 1-, 2-, and 3-most correlated donors, Corr(24), Corr(47) and Dist(68). These three graphs of SMG(25), SGM(47) and SGM(65) are selected so that it makes a fair comparison among the three methods as they all have a similar graph complexity. Validation errors associated with these three different levels of sparsity are represented in Figure 1 (b) by the three red, green, and magenta bars for the graphs of $G_{sgm}$, $G_{dist}$, and $G_{corr}$, respectively. Figure 1 (b) shows that the test errors (Equation (38)) are the lowest for the inferred daily streamflow time series using the $G_{sgm}$ graphs from the SGM algorithm, and are the highest based on the distance-based approach for all three cases with the donor gauges of 1, 2, and 3. The test errors for the inferred results using the pair-wise correlation-based approach are between the two for all three cases.

To test the statistical significance of these results shown in Figure 1 (b), procedures described to infer the streamflow time series were repeated 30 times with random selection of the records for the training and validation sets (keeping the test data set fixed). Running 6 single tailed t-tests using a significance level of 0.05, and a null hypothesis that the mean test error for the SGM graphs is equal to the Dist or Corr graphs (for the cases of 1-, 2-, and 3-donors, respectively), the null hypothesis was rejected in all cases (p-value < 0.0001), and the alternative hypothesis was accepted. That is, the mean error with the test data set for SGM(25) is significantly lower than that for Dist(24) and Corr(24); the mean error for SGM(47) is significantly lower than that for Dist(43) and Corr(47); and the mean error for SGM(65) is significantly lower than that for Dist(65) and Corr(68). In other words, the results obtained using our new method of the SGM algorithm are significantly better than those of using either the least distance-based or the maximum correlation-based approaches. Figure 1 (c) shows the relationships between the mean test error and the number of training days used for the least distance (green), maximum correlation (magenta) and SGM algorithm (red), respectively. The mean test error is the average test error with 1-, 2-, and 3-donor gauges or their equivalent counterparts in the SGM case for each method. The length of the training set varies from 45 days to 3650 days (ten years). The fifth point in each curve in Figure 1 (c) corresponds to 730 days (about 2 years). Figure 1 (c) shows that 2 years of training data are almost as good as the full range of 10 years. This result is important because for the case of ungauged basins, it is possible to place a temporary gauge station to collect data for about 2 years and then use the collected data to train the algorithms presented in this work to infer the streamflow time series for that specific ungauged location in the future as long as no dramatic environment change occurs for the study region.

Figure 2 shows how each of the individual graphs look like using the SGM, least distance (Dist), and maximum correlation (Corr) approaches. For the latter two approaches, the three commonly used scenarios with 1-, 2-, and 3-donors per target site are illustrated. The graphs for a single donor are Dist(24) and Corr(24) with their equivalent counterpart of SGM(25) from the SGM algorithm. For two donors they are Dist(43) and Corr(47), and their counterpart of SGM(47).

Finally, for three donors they are Dist(65) and Corr(68), and their counterpart of SGM(65). The graphs in Figure 2 with green edges are for the distance-based approach (Dist), magenta edges for the correlation-based approach (Corr) and red edges for the SGM approach. It can be seen that the graphs associated with each of the three approaches are not the same although some features in their graphic structures are similar. From Figure 1 (b) and the hypothesis testing results, it is clear that the new SGM method is the best of the three. This is because our new method with the SGM algorithm accounts for the dependence structure in the entire streamflow network based on the concept of conditional independence, and employs the Glasso method to effectively extract such dependence structure through making the precision matrix sparse. Our results demonstrate that a good use of the conditional independence structure of the underlying streamflow network (i.e., use sparse precision matrix) is important and it outperforms the widely used pair-wise correlation-based method (i.e., Corr) which only directly uses the local correlation information. Comparing to the distance-based method, the correlation-based method is superior which is consistent with other results reported in the literature.

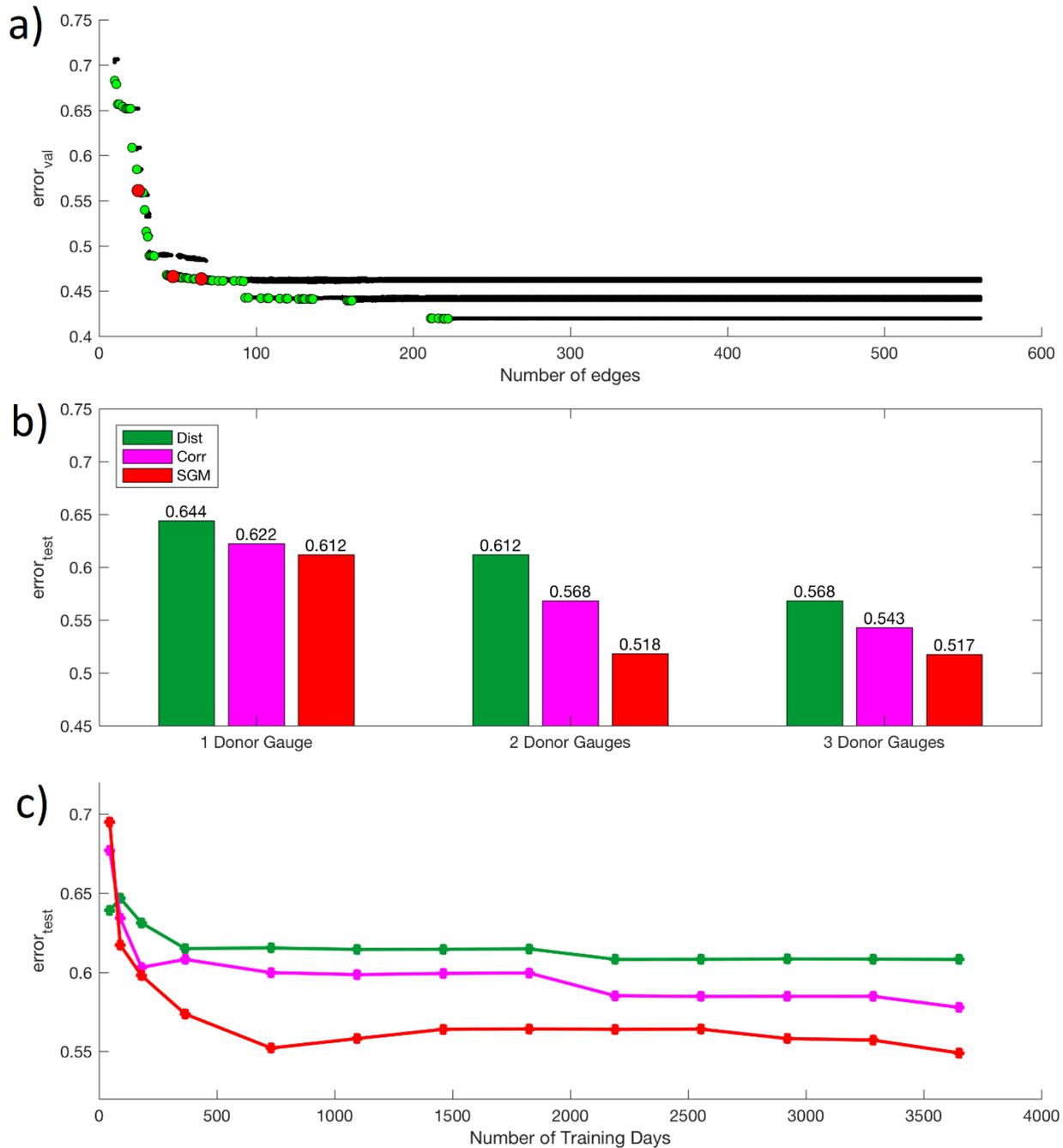

**Figure 1.** Result of running the SGM algorithm with the Ohio River Basin dataset. The training set is composed by a random selection of daily streamflow records between 1951 and 1970, while the validation set is composed by the remaining 50% for the same time span. The test data set is composed by the most recent 10 years of data (i.e. 1971-1980). **(a)** Validation error from the multi-objective optimization procedure of the SGM algorithm, between the observed and inferred daily streamflow time series vs the number of edges in the underlying graph (representing conditional independence assumptions between sites). The black dots represent sub-optimal (dominated) solutions. The Green dots represent the set of non-dominated (optimal) solutions. The red dots represent the graphs SGM(25), SGM(47) and SGM(65) with 25, 47 and 65 edges, respectively, chosen from the set of non-dominated solutions. **(b)** Comparison of the test error, between the SGM algorithm and the selection of donor gauges with the least distance (Dist) and maximum correlation (Corr) approaches, for 1, 2 and 3 donor sites. From left to right, comparison for one donor sites, the SGM(25) was selected to match the sparsity of Dist(24) and Corr(24). For two donor sites, graph SGM(47) was chosen to match Dist(43) and Corr(47). In the same way, SGM(65) is matched with Dist(65) and Corr(68). **(c)** Comparison of the mean test error as a function of the number of training days. Each point in (c) is calculated by averaging the values of the test error for 1-, 2-, and 3-donor gauges or their equivalent counterparts in the SGM case using a given number of training days, for 1, 2 and 3 donor gauges. The series for the least distance (Dist), maximum correlation (Corr) and SGM algorithm series are depicted in green, magenta and red, respectively. The size of the training set is 45, 90, 180, 365, 730, 1095, 1460, 1825, 2190, 2555, 2920, 3285, and 3650 days, respectively.

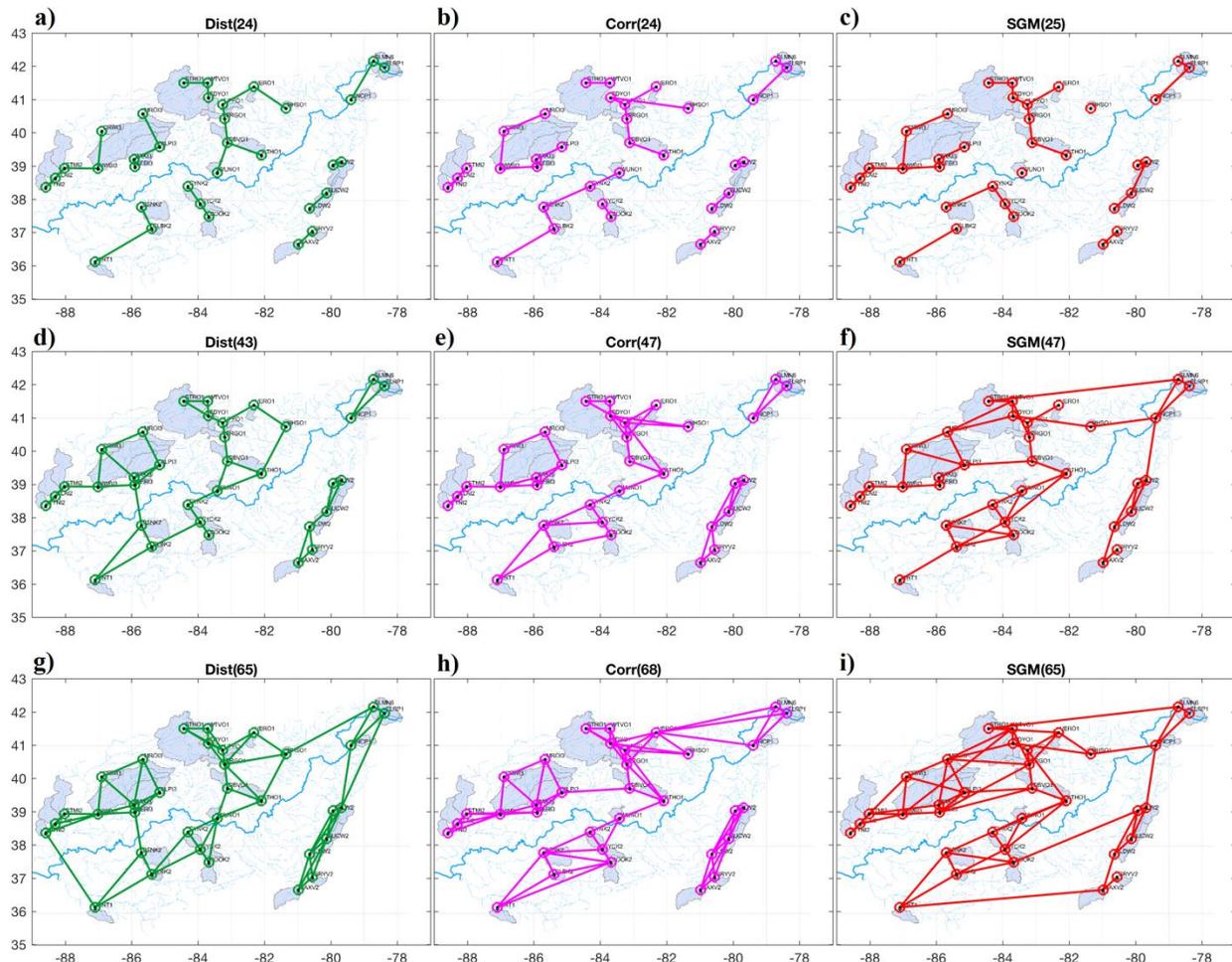

**Figure 2** – Comparison of the graphs generated by using the least distance (Dist) on left, the maximum correlation (Corr) in center column and the Selection of Graphical Model algorithm (SGM) on right, for 1 (on top), 2 (in middle row) and 3 (at bottom) donor sites, respectively. The graphs for 'Dist' are built by adding 1, 2 or 3 edge(s) from each site to the site(s) with the least distance for 1-, 2- and 3-donor sites, respectively. The graphs for 'Corr' are built by adding 1, 2 or 3 edge(s) from each site to the site(s) with the highest correlation for 1-, 2- and 3-donor sites, respectively. The number of edges for the 'Dist' and 'Corr' approaches is fixed depending on the number of donor sites used to build it, as opposed to the graphs from the SGM algorithm, where the number of edges can be selected from the set of non-dominated solutions. The SGM graphs are then selected to approximately match the sparsity of the 'Dist' and 'Corr' graphs for 1-, 2- and 3-donor sites, respectively. **(a)** Least distance graph with a single donor site, Dist(24) with 24 edges. **(b)** Maximum correlation graph with a single donor site, Corr(24) with 24 edges. **(c)** Selection of graphical model algorithm graph, SGM(25), with 25 edges. **(d)** Least distance graph with two donor sites, Dist(43) with 43 edges. **(e)** Maximum correlation graph with two donor sites, Corr(47) with 47 edges. **(f)** Selection of graphical model algorithm graph, SGM(47) with 47 edges. **(g)** Least distance graph with three donor sites, Dist(65) with 65 edges. **(h)** Maximum correlation graph with three donor sites, Corr(68) with 68 edges. **(i)** Selection of graphical model algorithm graph, SGM(65) with 65 edges.

**5.2 Removal of gauges with least loss of information**

The objective here is to remove several gauges from the gauge network with the least loss of information. For comparison of the three approaches, graphs for a single donor, Dist(24), Corr(24), and SGM(25), two donors, Dist(43), Corr(47), and SGM(47), and three donors, Dist(65), Corr(68) and SGM(65), are used. Each of the nine graphs shown in Figure 2 is served as an input to the Removal of Gauges (RG) algorithm described in section 3.7. Although the RG algorithm shows that there are $8 - 16$ gauges that can potentially be removed, only about $7 - 8$ gauges among them can be inferred with an NSE higher than 0.7.

Figure 3 shows the comparison results in which location of each gauge is the same as that in Figure 2. The removable gauges are represented by color-coded circles in Figure 3, indicating their corresponding inference accuracy, measured by an NSE value, when they are estimated by other gauges. An NSE value higher than or equal to 0.9 is depicted in blue; between 0.8 and 0.9 in green; between 0.7 and 0.8 in yellow; between 0.6 and 0.7 in orange; and below 0.6 in red. From Figure 3, one can clearly see that in general there are more gauges that can be removed from the SGM approach than those from the other two approaches. Furthermore, for the same gauges identified to be removable by all three methods, the accuracy achieved (or information lost) by the SGM method is the highest (least) as there are more blue, green and yellow circles combined in Figure 3 for the SGM method than for the other two methods.

Figure 4 (a) shows that the graph score for the single donor case is higher for the SGM approach than for the other two approaches. For the case with two donors, the SGM approach significantly outperforms the other two methods. The case for three donors is less clear, as there is a tie between the Corr and the SGM methods. A closer examination of the results indicates that the chosen threshold ($\Gamma = 0.7$) is the culprit. For the three donors case, SGM(65) allows the removal of 11 gauges, but only seven with an NSE greater than 0.7 while two have NSE between 0.6 and 0.7 and the other two have NSE below 0.6. On the other hand, Corr(68) allows the removal of only 8 gauges with 7 of them having an NSE greater than 0.7 and the eighth one below 0.6. Because the graph score only takes into account of the gauges with an NSE higher than $\Gamma$, the two additional gauges with NSE=0.65 in the SGM method were ignored, resulting in the same value in the bar plot in Figure 4 (a) for the 3 donor gauge case, while in fact they are different as the SGM(65) approach allows for a removal of two additional gauges with just a slightly lower NSE value than the prescribed threshold. These two gauges are highlighted in orange in Figure 3 (i). In addition, one can see that among the seven common removable gauges, the SGM method has two gauges in blue in Figure 3 (i) while the Corr method only has one gauge in blue and the other top north-east one in green in Figure 3 (h). Figure 4 (b) shows that the mean graph score is higher for the SGM method than those for the two other approaches, with the Corr method in second place and the Dist method in third. Figure 4 (c) shows a related but slightly different measure to that of Figure 4 (a) in assessing the quality of the inference results for the streamflow time series estimated by the three methods of SGM, Dist and Corr. In Figure 4 (c) the mean is taken from the 8 removable gauges with the highest NSE for each of the donor scenarios. The SGM has a higher mean among the top 8 removable gauges, for 1-, 2- or 3-donors, than the other two approaches. Figure 4 (d) shows the mean NSE for each method presented in Figure 4 (c). Clearly, the mean NSE for the top 8 removable gauges is higher for the SGM method than those for the other two methods. In order to validate the statistical significance of these results shown in Figure 4, a similar procedure as the one done for Figure 1(b) was performed, using a set of 6 single tailed t-tests. They are: (1) SGM(25) vs Dist(24), (2) SGM(25) vs Corr(24), (3) SGM(47) vs Dist(43), (4) SGM(47) vs Corr(47), (5) SGM(65) vs Dist(65), and (6) SGM(65) vs Corr(68) for Figures 4 (a) and (c). For Figures 4 (b) and (d) a set of 2 single tailed t-tests were used. They are: (1) SGM vs Dist, and (2) SGM vs Corr. It was concluded that the results shown on Figure 4 (b), (c) and (d) are statistically significant at a significance level of 0.05. The results shown on Figure 4 (a) are less conclusive as expected, however the results were statistically significant in 4 out of 6 of the tests. The exceptions were cases (2) SGM(25) vs Corr(24), and (6) SGM(65) vs Corr(68) due to the prescribed threshold of 0.7 discussed above.

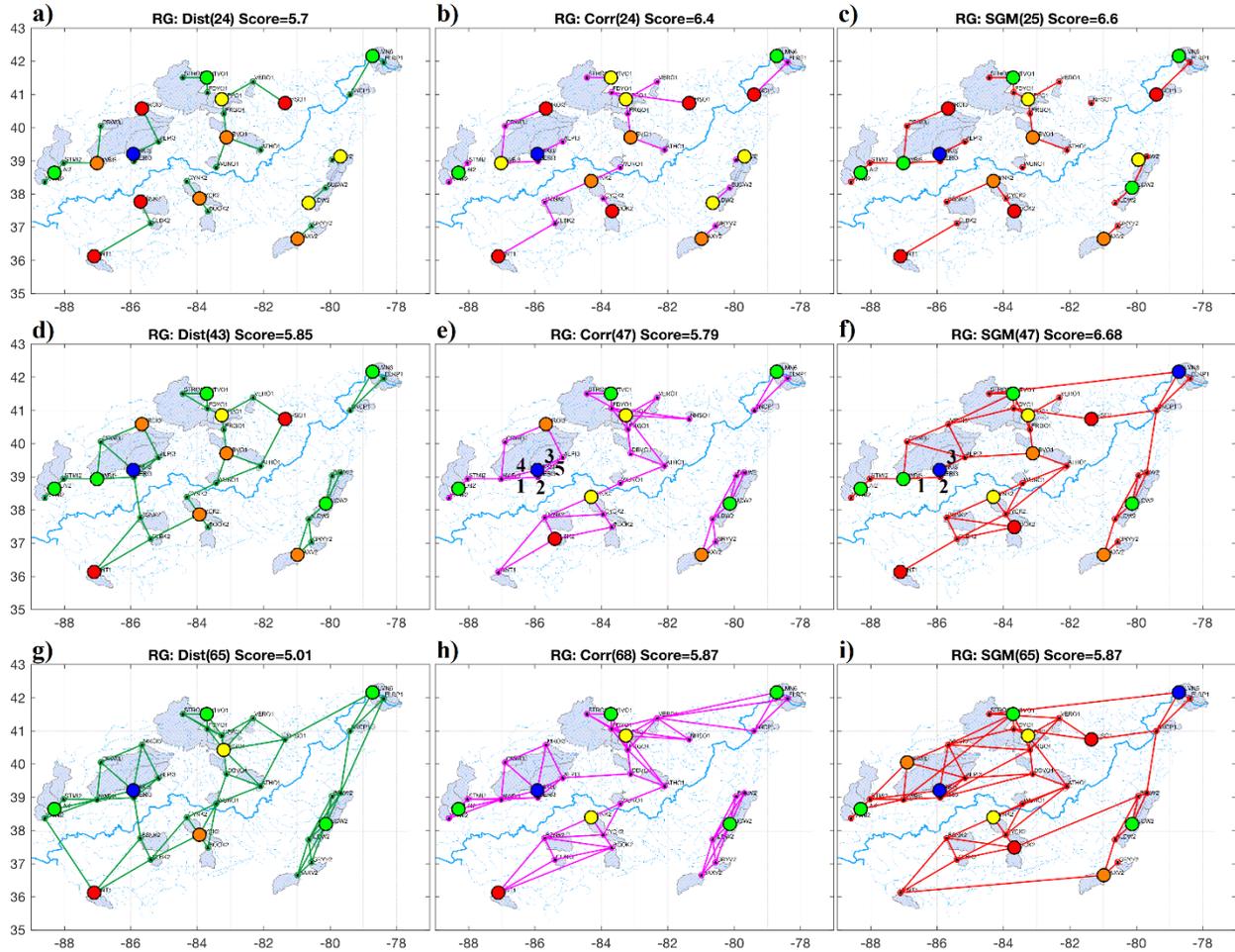

**Figure 3** – Comparison of the observed and inferred daily streamflow time series, in the *test* set (records between 1971 and 1980), for removable gauges estimated by the Removal of Gauges (RG) algorithm using the graphs in Figure 2 as inputs. Least distance (Dist) on the left, the maximum correlation (Corr) in the middle and the Selection of Graphical Model algorithm (SGM) on the right, for donor sites of 1 (on the top), 2 (in the middle row) and 3 (at the bottom), respectively. Note that for the SGM case, the number of donor sites are not fixed but automatically determined. The target sites chosen by the RG algorithm are highlighted in blue for Nash Sutcliffe efficiency (NSE) greater than or equal to 0.9, in green for NSE between 0.8 and 0.9, in yellow for NSE between 0.7 and 0.6, in orange for NSE between 0.6 and 0.7, and in red for NSE < 0.6. The graph *score* is the sum of the NSE for the subset of the inferred target sites with NSE > 0.7. The meaning of each plot is: **(a)** Least distance graph of Dist(24) for a single donor site with 24 edges and a score of 5.7. **(b)** Maximum correlation graph of Corr(24) for a single donor site with 24 edges and a score of 6.4. **(c)** SGM graph of SGM(25) with 25 edges whose sparsity is similar to the single donor case of graphs Dist(24) and Corr(24). It has a score of 6.6. **(d)** Least distance graph of Dist(43) for two donor sites with 43 edges and a score of 5.85. **(e)** Maximum correlation graph of Corr(47) for two donor sites with 47 edges and a score of 5.79. Five edges are marked in the plot as: 1 (NWBI3-SERI3), 2 (BAKI3-SERI3), 3 (ALPI3-BAKI3), 4 (NWBI3-BAKI3), and 5 (ALPI3-SERI3) **(f)** SGM graph of SGM(47) with 47 edges and a score of 6.68. Three edges are marked in the plot as: 1 (NWBI3-SERI3), 2 (BAKI3-SERI3), and 3 (ALPI3-BAKI3). **(g)** Least distance graph of Dist(65) for three donor sites with 65

edges and a score of 5.01. (**h**) Maximum correlation graph of Corr(68) for three donor sites with 68 edges and a score of 5.87. (**i**) SGM graph of SGM(65) with 65 edges and a score of 5.87.

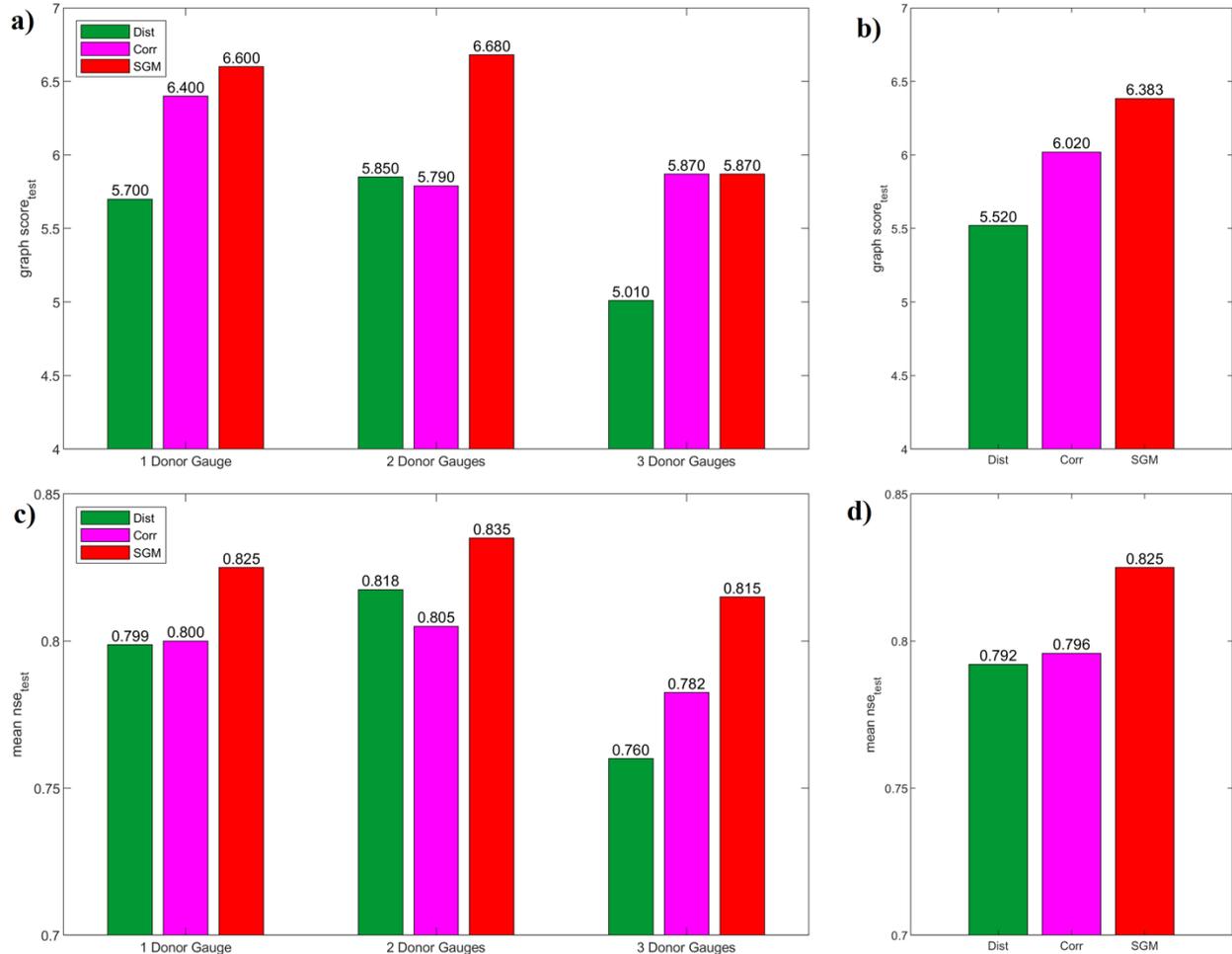

**Figure 4** - Comparison of the inference accuracy on the removable gauges with the RG algorithm applied to graphs corresponding to the SGM, the least distance criterion (Dist) and the maximum correlation (Corr) methods. The meaning of each plot is: (**a**) Graph score (Equation (41)) for each of the individual graphs for the testing data set. (**b**) The mean graph score of the 1-, 2- and 3-donors for each method based on values shown in (a). (**c**) Mean NSE of the eight removable gauges with the highest NSE. (**d**) The mean NAE of the 1-, 2- and 3-donors for each method based on values shown in (c).

In general, the pair-wise correlation-based approach (Corr) is more accurate than the distance-based approach (Dist). But the former requires more and better data to establish the correlations. From the correlation perspective, the SGM method is more similar to the Corr approach than to the Dist approach which only depends on the geographical locations of the sites. However, there is a fundamental difference between the SGM method and the Corr method. The widely used Corr method uses the pair-wise correlation to determine the edges between the sites. As a consequence, sites that have a decent correlation with other sites will end up with a relatively large number of edges associated with them. But some of these edges are redundant. On the other hand, the new SGM method takes the advantage of the conditional independence condition between sites as oppose to the pair-wise correlation used by the Corr approach. Therefore, the SGM method reduces the amount of redundant edges between sites and only connects a subset of these sites. In addition, the SGM method uses the precision matrix instead of the correlation matrix which makes it easier to extract the dependence structure among the sites within the entire network. These good characteristics associated with the SGM method, in turn, depict a simpler and more accurate dependence structure of the underlying gauge network for the study region. In practice, this simpler and better gauge network increases the number of sites that can be inferred without a significant loss in accuracy. That is, the graph from the SGM method can distribute the "correlated flow" in a more efficient way. Our results in Figures 3 and 4 have shown that indeed the accuracy of the inferred streamflow time series is improved and that the number of potentially removable sites is also increased compared to the Corr approach.

One clear example of the difference between the marginal (Corr) and the conditional (SGM) correlation methods is given by the relationship identified between the sites ALPI3, BAKI3, NWBI3 and SERI3 shown in Figures 3 (e) and 3 (f). BAKI3 with a catchment area of 4421 $Km^2$ is a sub-basin of SERI3 with a catchment area of 6063 $Km^2$ along the main channel. Therefore, the catchment area of BAKI3 accounts for 73% of the catchment area of SERI3 and the correlation between them is the highest among the sites considered in the study area. The edge between them is present in all of the nine graphs shown in Figure 3. The sites BAKI3 and SERI3 are also highly correlated to the sites ALPI3 and NWBI3. Figure 3 (e) shows the graph for Corr(47), with five edges: 1 (NWBI3-SERI3), 2 (BAKI3-SERI3), 3 (ALPI3-BAKI3), 4 (NWBI3-BAKI3), and 5 (ALPI3-SERI3). Figure 3 (f) shows the graph for SGM(47) with only three edges which are the same edges as shown in Corr(47), but having the following two edges, NWBI3-BAKI3 and ALPI3-SERI3, dropped. It is safe for SGM(47) to drop these two edges as NWBI3 is conditionally independent to BAKI3 given SERI3, and ALPI3 is conditionally independent of SERI3 given BAKI3. Figures 4 (a) and (c) show that among the nine graphs shown in Figure 3 the graph with the best trade-off between model complexity and accuracy is SGM(47).

Figure 5 shows the comparison between the observed and inferred daily streamflow time series based on the testing set for the eight streamflow gauges with the highest NSE, when SGM(47), shown in Figure 3 (f), is chosen as the underlying graphical model.

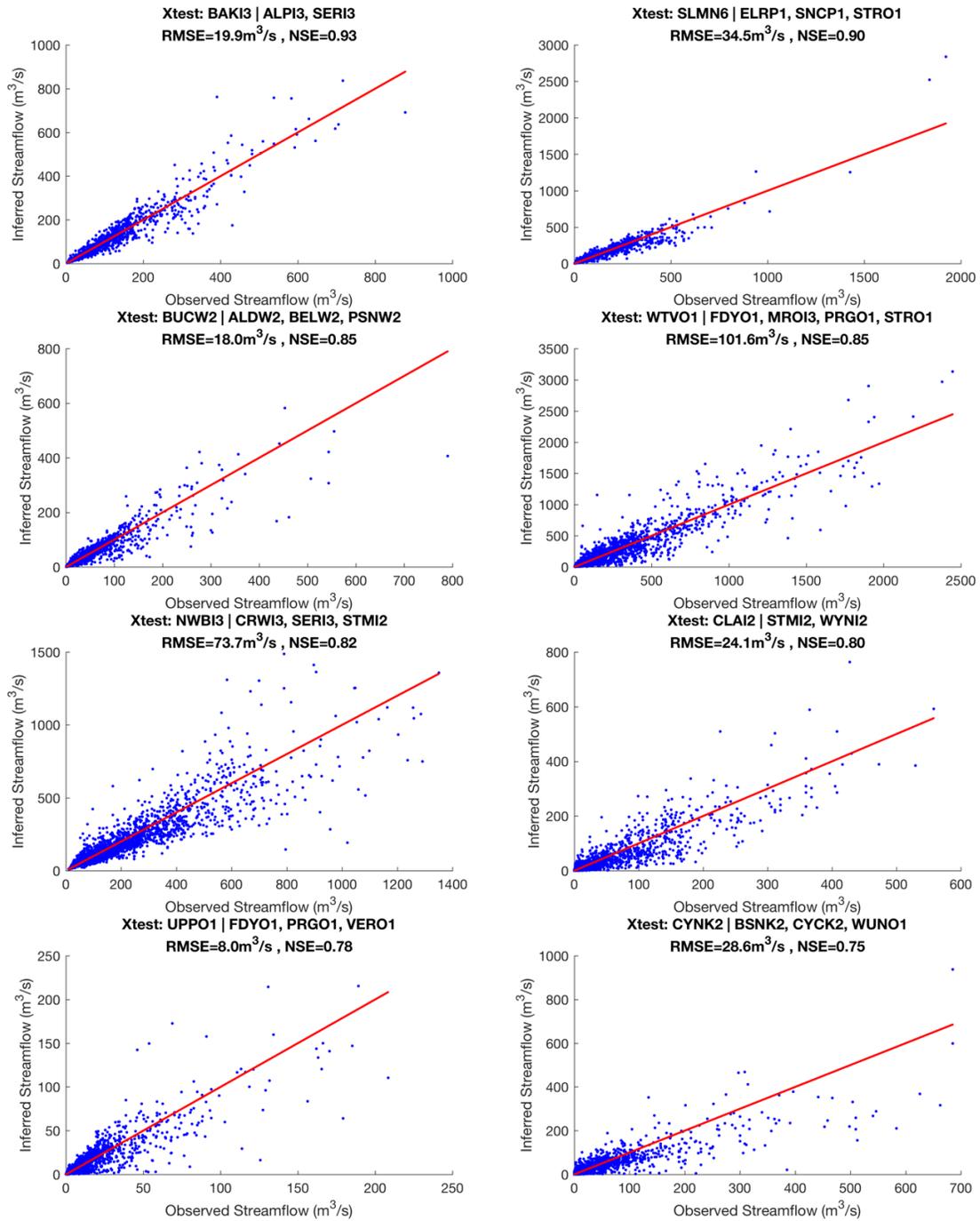

**Figure 5.** Scatter plots between the observed and inferred daily streamflow time series over the test period of 1971-1980 (i.e., test data set). Each plot represents one of the eight gauges with the highest NSE values among the removable gauges shown in Figure 3. The RG algorithm is used in combination with the SGM(47) graph to identify the gauges to be removed. The MLR with Equations (34) and (35) is used to infer the daily streamflow shown in the plots. The root mean squared error (*RMSE*) and the *NSE* are shown for each gauge over the inferred period of 10 years. At the top of each plot, the name of the removed gauge is indicated on the left side of the divide line "|", and the names of gauges used to infer the streamflow of the removed gauge are indicated on the right side of the divide line "|".

For each of the 34 gauges, their corresponding watersheds were delineated using a Geographical Information System (GIS) to facilitate our understanding of the identified connections and isolated gauges based on the SGM method. Figure 6 (a) shows the elevation (NED: National Elevation Dataset), (b) slope (derived from elevation data), (c) soil type (Hybrid STATSGO/FAO Soil Texture) and (d) land cover (MRLC: Multi-Resolution Land Characteristics Consortium) along with the selected non-dominated graph SGM(25) obtained with a GIS tool and the corresponding cited data sets.

Using SGM(25), two sites are isolated. NHSO1 and WUNO1. Isolated sites should be maintained as much as possible to avoid loss of important regional information. Less sparse graphs, such as SGM(47) and SGM(65), can still have some marginal benefit from having some edges to those sites. NHSO1 has a significantly different land use comparing to other watersheds in the study region. For NHSO1, more than 50% of its drainage area is developed while others have less than 20% as developed. Thus, the hydrological response of this watershed to precipitation events is very different from other watersheds. In the case of WUNO1, its isolation in SGM(25) appears to be related to a combination of its geographic location, different land use from its neighboring watersheds, and its proximity to the main channel of the Ohio River. This last factor seems to be a natural separator of it. There are no edges crossing the Ohio River on the selected sparse graph, SMG(25) with 25 edges, shown in Figure 2 (c).

In general, the factors that impact the connections (i.e., conditional correlations) between gauges are complex and it is the integrated effect (e.g., the streamflow in this case) that determines the conditional correlations between the gauges. The first-order factors that contribute to the generation of streamflow in the study area seem to be the elevation, the slope and the catchment area. There is a relatively high correlation between the specific discharge (i.e., streamflow divided by the catchment area) and the elevation (0.79), and between the specific discharge and the slope (0.76). The land cover also plays an important role, as the edges in SGM(25) are usually between sites with the same land cover class as shown in Figure 6 (d).

Results here have demonstrated again that it can be difficult to just use relatively simple and explicit functions to relate streamflow to different factors such as land cover, slope, soil type, drainage size in identifying their connections for complicated situations like this study case. Such a point has also been illustrated in the literature (e.g., Parada and Liang, 2010). On the other hand, these factors can sometimes help us understand why certain links exist while others do not. For example, the land cover types, elevation, and slopes appear to play more important roles than the soil type in this study region. It is worth pointing out that gauges are sometimes connected even if the correlations between them are not very high. They are connected simply because there are no other available gauges nearby with acceptably higher (conditional) correlations. In summary, the chosen graph SGM(25) does not have any edges crossing the Ohio River; there exist two gauges isolated from the rest, those gauges are geographically far from other gauges and one of them has a significantly different land use category distribution with more than 50% of its area being developed. Most of the area of the Ohio River basin belongs to the same soil type category and therefore, the soil type does not appear to contribute to the identification of the hydrologic similarity between sub-basins in this study case. On the other hand, most of the edges on the selected underlying graph SGM(25) are between watersheds with the same land use category. These results suggest that in the Ohio River basin, the land use is an important factor for the hydrologic similarity among the sub-basins.

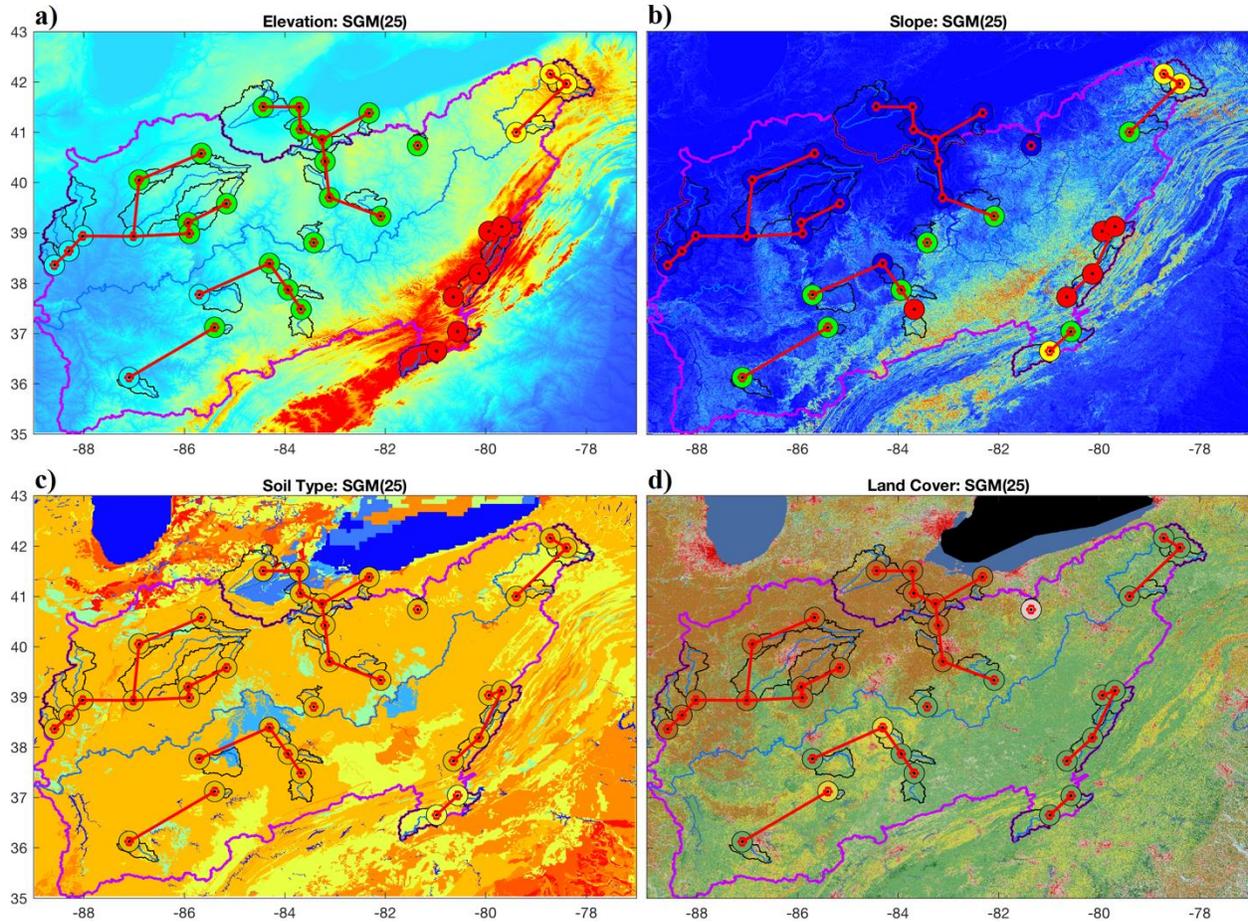

**Figure 6.** Spatial distributions of the elevation, slope, soil type, and land cover over the study region with graph SGM(25). (a) Elevation map showing a cluster of 4 categories indicated by a filled circle of cyan, green, yellow and red respectively. These different colors represent, respectively, "very low", "low", "high" and "very high" elevations based on the mean elevation of their corresponding watersheds. (b) Slope map showing a cluster of 4 categories indicated by a filled circle of blue, green, yellow and red respectively. The colors represent, respectively, "very low", "low", "high" and "very high" slope based on the mean slope of their corresponding watersheds. (c) Soil type map showing a cluster of 2 categories indicated by a filled circle of dark yellow and light yellow, respectively. These two colors represent, respectively, the "silt loam" and "loam" soil types. (d) Land cover map showing a cluster of 4 categories indicated by a filled circle of pink, green, yellow and brown, respectively. These different colors represent, respectively, the "developed, open space", "deciduous forest", "pasture/hay" and "cultivated crops" land cover types.

## 6. Conclusions

In this study, we proposed a novel method, *Selection of Graphical Model* (SGM) algorithm, to select the set of multiple donor gauges for each inactive gauge (i.e., target gauge) to extend/infer its daily streamflow time series. This method generates a series of graphs that represent the set of potential donors for each site. The graphs generated by the SGM algorithm allow more accurate estimation of daily streamflow time series than other commonly used approaches based on the distance between sites (Dist) and the pair-wise correlation (Corr) between the streamflow time series. The main idea of our new SGM method is to take the advantage of the conditional independence structure encoded by an undirected graphical model known as Gaussian Graphical model, represented by the precision (i.e., covariance inverse) matrix. The SGM method selects multiple donor gauges by imposing sparsity to the precision matrix via the Graphical Lasso algorithm. The two parameters, the L1 norm regularization and a truncation threshold, in the SGM algorithm are determined by a multi-objective optimization procedure that minimizes both the number of edges of the underlying graph and the error in the validation set to achieve a balance between the sparsity and connectivity/complexity for each graph. The resulting graphs from the non-dominated solution encode the set of donor gauges that are then used for the inference of the daily streamflow for each target gauge. We have illustrated in this study that for the gauge network composing of 34 daily streamflow gauges in the Ohio River basin, the graph with 47 edges selected based on our SGM algorithm has a good trade-off/balance between network sparsity and the estimation error. With our RG algorithm, a set of gauges can be removed from the hydrometric network with the least loss of information. In this study (e.g. Figure 3 and Figure 4), we have demonstrated that eight out of 34 (25%) gauges can potentially be removed (NSE >= 0.75), and that from them, a group of six (18%) gauges can be inferred with relatively high accuracy (NSE >= 0.8) using the donors identified by the SGM method. In addition, for all three cases with 1-, 2-, and 3-donors, the new SGM method outperforms the commonly used the least distance (Dist) approach and the maximum correlation (Corr) approach (see Figure 1 (b), Figure (3) and Figure (4)) in inferring the daily streamflows on the removable gauges. Sensitivity on the length of daily data required for achieving a stable SGM graph was investigated. Our results (see Figure 1 (c)) show that a length of about 2-year daily data is needed. For applications to ungauged basins, a feasible way is to install a temporary streamflow gauge to collect data for about two years. Such data would then be used with data from other gauges in the network to obtain the SGM graph. Then, daily streamflow time series for the ungauged basin can be inferred using the methodology described in this study as long as there is no dramatic environment change over the study region. We will test our new method and compare its results to other methods for inferring the daily streamflow data at the ungauged basins, and report the comparison results at its due course in the future.

Depending on the number of gauges needed for removal, a balance between the inference accuracy and the gauge removal numbers can be achieved. In general, the sparser the graphs are, the more gauges one can remove. On the other hand, our study also demonstrates that the complete graph (i.e., with 561 edges) is not included in the set of non-dominated solutions, indicating that having more donor gauges does not necessarily achieve optimum results due to significantly more noises (inconsistency) introduced by the data and the inclusion of redundant edges. Therefore, not only can a suitable sparse graph achieve better inferring results through finding the most essential correlations, but also it is more practical because it requires a small but most relevant number of donor gauges in inferring the streamflow for the inactive gauges and a fewer observations to establish the relationship through the data training process. Furthermore, a graph with a fewer edges can reduce overfitting. Our method has two limitations. First, it requires a historical record of two years or more to characterize the relationships between the target and donor gauges. Second, the probability distribution of the daily streamflow should be approximated well by a log-normal distribution so that the log-transformed variable distributes normally. This second limitation, however, can be easily overcome through a common distribution transformation method if the log-normal assumption does not hold. In this study, the inference stage was performed with an ordinary least squares MLR approach due to its simplicity, although other approaches can also be used once a set of donor gauges is identified.

The most computationally expensive part of our new method is the SGM algorithm, as it relies on calling the Graphical Lasso method multiple times to find the optimal combination of the regularization and truncation parameters. However, this complexity can be abstracted by calling efficient routines such as the glasso Matlab package (glasso) and also the GLASSOFAST (Sustik & Calderhead, 2012) package which is a faster and more recent implementation of the Graphical Lasso algorithm. In addition, in this work, we performed a thoroughly search for the regularization parameter between 0 and 1, but it was determined that the best range was between 0.01 and 0.1, and that using just 10 values was almost as good as using 30 values in between. Also, we performed in this study an almost exhaustive search for the truncation parameter to go from a very sparse graph with only 10 edges to a full graph with 561 edges (for a graph of 34 nodes), but we demonstrated that sparse graphs, under 65 edges, achieve better overall results, allowing the accurate inference of multiple sites with relatively high accuracy (NSE $>= 0.8$).

In this work, only contemporaneous daily streamflow records are considered. The methods explained here can be adapted to include lagged records for a finite set of days. However, for the sake of simplicity such approach was not followed. Related work (Farmer, 2016; Skøien &

Blöschl, 2007) found only marginal improvements when considering streamflow travel times into geostatistical analysis.

**Acknowledgments and Data**

This work was partially supported by the William Kepler Whiteford Professorship from the University of Pittsburgh. We thank the Center for Research Computing at the University of Pittsburgh in providing computing resources for this work.

For this work, German A. Villalba implemented the research ideas, designed algorithms, performed experiments, conducted analysis, and co-wrote the manuscript. Xu Liang conceived the research ideas, designed experiments, supervised the investigation, and co-wrote and finalized the manuscript. Yao Liang conceived the research ideas and co-wrote the manuscript.

We would like to thank the scientists who worked at the U.S. Geological Survey for their streamflow data (https://waterdata.usgs.gov/nwis) and digital elevation model (https://viewer.nationalmap.gov/basic/?basemap=b1&category=ned), the National Center for Atmospheric Research for the soil type and vegetation data (https://ral.ucar.edu/solutions/products/noah-multiparameterization-land-surface-model-noah-mp-lsm), and the Multi-Resolution Land Characteristic Consortium (https://www.mrlc.gov/) for their land-cover data. All of the data used in this study can be freely obtained from their respective websites as they are all publicly available. The Matlab codes for running our new algorithms and the datasets used to generate the tables and figures in this paper can be accessed from DOI: 10.5281/zenodo.3634206.